\documentclass[
reprint,
superscriptaddress,
nature
]{revtex4-2}

\usepackage{svg}
\usepackage{xcolor}
\usepackage{graphicx}
\usepackage{dcolumn}
\usepackage{bm}
\usepackage{amsmath}
\usepackage{hyperref}
\usepackage{physics}
\usepackage{sidecap}
\usepackage{floatrow}
\usepackage{xr}
\usepackage{booktabs}
\usepackage{tabularx}
\usepackage{makecell}
\usepackage{amsfonts}
\usepackage{array}
\usepackage{tabularray}
\usepackage{siunitx}

\begin{document}

\preprint{}

\date{\today}

\title{Quantum coherent transceivers toward Holevo-limited communications}

\author{Volkan Gurses}
\thanks{Corresponding author: gurses@caltech.edu}
\affiliation{Division of Engineering and Applied Science, California Institute of Technology, Pasadena, CA, USA}
\affiliation{Division of Physics, Mathematics and Astronomy, California Institute of Technology, Pasadena, CA, USA}
\author{Suraj Samaga}
\affiliation{Division of Engineering and Applied Science, California Institute of Technology, Pasadena, CA, USA}
\author{Elianna Kondylis}
\affiliation{Division of Engineering and Applied Science, California Institute of Technology, Pasadena, CA, USA}
\author{Ali Hajimiri}
\affiliation{Division of Engineering and Applied Science, California Institute of Technology, Pasadena, CA, USA}

\begin{abstract}
The Holevo limit bounds the channel capacity of a communication channel in which information is encoded in quantum states in a Hilbert space at the transmitter and decoded using quantum measurements at the receiver. Saturating the Holevo limit requires quantum-limited transceivers that either generate quantum states of light or employ quantum-limited measurements. Here, we demonstrate an integrated photonic-electronic quantum-limited coherent receiver (QRX) achieving 14.0~dB shot noise clearance (SNC), 520~$\mu$W knee power, 2.57~GHz 3-dB bandwidth, 3.50~GHz shot-noise-limited bandwidth, and 90.2~dB common-mode rejection ratio ($\mathrm{CMRR}$). We scale this design to a 32-channel QRX array with median 26.6~dB $\mathrm{SNC}$, and automatic $\mathrm{CMRR}$ correction yielding a median 76.8~dB $\mathrm{CMRR}$ at minimum. Using the integrated QRX and fiber-optic transmitter, we measure $0.15\pm0.01$~dB of squeezing below the shot noise limit, limited by off-chip losses. We propose a squeezed light communication scheme that can surpass the Shannon limit, with a path toward the Holevo limit.
\end{abstract}

\maketitle

\section{Introduction}

\begin{figure*}
    \centering
    \includegraphics[width=\linewidth]{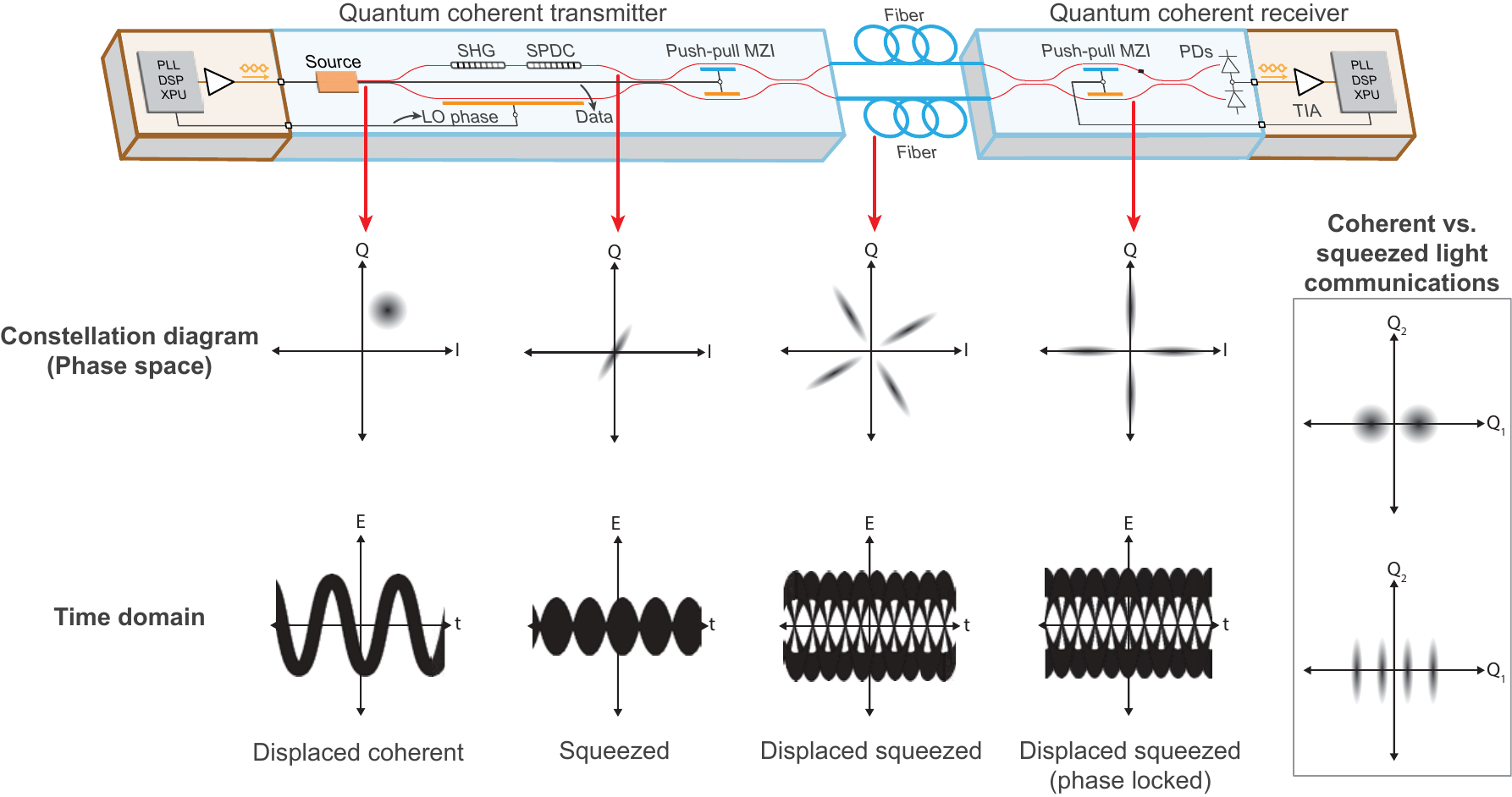}
    \caption{\textbf{Squeezed light communications with quantum coherent transceivers.} Constellation diagram (phase space) and time domain representations at each stage of the quantum coherent transceiver. (1, Displaced coherent) The laser outputs a coherent state, a Gaussian symbol displaced from the origin, corresponding to a sinusoidal carrier with uniform noise for each quadrature angle and is split into a signal path (top) and a local oscillator (LO) path. (2, Squeezed vacuum) The signal light pumps a second harmonic generation (SHG) process, upconverting the field, which then pumps a spontaneous parametric downconversion (SPDC) process that produces a quadrature-squeezed vacuum state centered at the origin, represented by a tilted ellipse. In the time domain, this is a zero-mean field with periodically amplitude-modulated noise. (3, Displaced squeezed vacuum) Interfering the squeezed vacuum with LO displaces it and a modulator phase modulates the state to one of four quadrature phase shift keying (QPSK) constellation points, preserving the elliptical noise contour, visible as squeezed noise riding on a sinusoidal carrier. (4, Phase locked displaced squeezed vacuum) At the receiver, the LO is phase-locked to the squeezed quadrature, aligning detection to the minimum-noise axis and recovering the squeezing sensitivity advantage below the shot noise limit. Inset showing the increase in channel capacity with squeezed light by packing more symbols in the same range in the constellation diagram.}
    \label{fig:fig1}
\end{figure*}
The ultimate channel capacity of a communication channel is set by the Holevo limit, which bounds the accessible mutual information per channel use when quantum measurements are employed at the receiver. Coherent states can saturate this bound when used with joint detection receivers. While this regime has been studied extensively \cite{Giovannetti2004LossyBosonicCapacity,Shapiro2009QuantumTheoryOpticalComm,Guha2011StructuredReceivers,GuhaDuttonShapiro2011JDR} and demonstrated experimentally \cite{Chen2012ConditionalNulling,Cui2025GreenMachine}, the use of non-classical states for communications has remained more limited. In particular, squeezed light can improve communication performance by redistributing quantum noise between conjugate quadratures, reducing fluctuations along one quadrature at the expense of the other.

Coherent detection resolves both quadratures of the optical field by interfering the incoming signal with a phase-locked local oscillator onto a balanced photodiode pair. Unlike direct detection which measures only intensity, this phase sensitivity gives access to the full quadrature statistics of the incoming field including the sub-shot-noise fluctuations of a squeezed state. Beyond classical communications and sensing, quantum-limited coherent receivers can therefore enable quadrature measurements of non-classical light, provided the receiver noise floor is dominated by shot noise rather than electronic noise across the operating bandwidth. Non-classical light encompasses states of light with properties that can be explained only with the quantization of the electromagnetic field. These non-classical properties can be leveraged with quantum(-limited) coherent transceivers to enable, for instance, optical links with sub-shot-noise-limited sensitivity \cite{Gurses2022, Gurses2024}. This improvement in sensitivity can be used to increase the signal-to-noise ratio ($\mathrm{SNR}$) in optical sensing and increase the channel capacity in optical communications \cite{Banaszek2020, SlusherYurke1990SqueezedLightCoherentCommunications}. High-bandwidth quadrature measurements of quantum states of light also enable continuous-variable quantum key distribution, quantum random number generation, quantum state engineering, and photonic quantum computing. Therefore, the realization of quantum coherent transceivers with large-scale photonics and electronics has the prospect of impacting several applications.

An exemplary communications scheme with squeezed light is illustrated in Fig. \ref{fig:fig1}. It utilizes integrated quantum coherent transmitters and receiver to encode and decode data using squeezed light as a carrier. The quantum coherent transmitter begins with a displaced coherent state from the laser source, split to a signal and local oscillator (LO) path, generates a squeezed vacuum state by a cascaded second harmonic generation (SHG) and then spontaneous parametric downconversion (SPDC) process, interferes it with LO to generate a displaced squeezed state and then coherently modulates the displaced squeezed state to encode symbols, shown here for a quadrature phase shift keying (QPSK) constellation. In phase space, this displacement preserves the anisotropic noise ellipse of the squeezed state, so that the squeezed quadrature can be aligned with the decision axis separating neighboring symbols. In the time domain, the same process appears as a carrier whose fluctuations are selectively suppressed in the measured quadrature rather than distributed uniformly. At the receiver, the local oscillator is phase locked to the squeezed quadrature so that coherent detection projects the incoming field onto the minimum-noise axis and recovers the squeezing advantage below the shot-noise limit. This quadrature-selective noise suppression increases the effective distinguishability of neighboring symbols at a fixed mean photon number and can therefore increase the achievable channel capacity or data rate for the same optical power \cite{YuenShapiro1978TwoPhotonCoherentStates,SlusherYurke1990SqueezedLightCoherentCommunications,ChesiOlivaresParis2018SqueezingEnhancedPSK,FanizzaRosatiSkotiniotisCalsamigliaGiovannetti2021SqueezingEnhancedCommunication}. Realizing this advantage in practice requires highly integrated coherent transceivers that can generate, transport, and detect quantum light coherently and at scale.

While integration of coherent transceivers for classical communications and sensing has reached maturity and commercialization, coherent transceivers with quantum-limited sensitivity that can probe non-classical states of light have only recently been demonstrated \cite{Gurses2023,Tasker2021,Tasker2024}. Silicon photonics offers a scalable platform to integrate quantum coherent transceivers with high-performance electronics for large-scale quantum optoelectronic circuits, enabling a high level of parallelism in a compact form factor and close interfacing between quantum photonics and electronics. Optimization of the form factor with close packaging of photonics and electronics is critical to ensure minimal parasitics between the photonic integrated circuit (PIC) and electronic integrated circuit (EIC). Co-design of photonics and electronics is also critical to minimize optical loss, electronic noise, and maximize local oscillator (LO) power and common-mode rejection ratio ($\mathrm{CMRR}$), to ensure that the receiver operates with a noise floor dominated by the shot noise of the signal field.

In this work, we outline the theory and design guide for quantum coherent transceivers that can transmit and receive non-classical light. We then demonstrate an integrated photonic-electronic quantum coherent receiver (QRX) along with a fiber-optic quantum coherent transmitter (QTX) that can generate and transmit squeezed light. The QRX has a shot noise clearance ($\mathrm{SNC}$) of 14.0 \unit{dB}, a knee power ($\mathrm{P_{knee}}$) of 520 $\mathrm{\mu W}$ with 3-\unit{dB} and shot-noise-limited bandwidths of 2.57 GHz and 3.50 GHz, respectively. The packaged system is compact with a total footprint of 2.7 $\mathrm{\times}$ 0.8 $\mathrm{mm^2}$. We also demonstrate large-scale integration of these receivers on chip by showing a 32-channel QRX array and demonstrating quantum-limited performance at scale. Lastly, we evaluate the capacity and energy efficiency gains achievable with squeezed light communications in the pursuit to push communications from the Shannon limit to the Holevo limit. Using the described coherent receiver model, we compare the SNR, channel capacity, and data rate, and energy per bit achievable with squeezed light against the Shannon and Holevo limits and show that squeezed light communications with coherent detection occupies an intermediate regime between the two. The capacity gain from squeezing over the Shannon limit comes at the cost of pump power overhead, and we identify the parametric gain coefficient $\mu$ of the nonlinear waveguide and the end-to-end detection efficiency $\eta$ as the two key parameters that govern whether this tradeoff is favorable.

\section{Results}
\subsection{Coherent receiver theory}

The background for the semi-classical and quantum treatments of coherent detection, including the second quantization of the electromagnetic field, is developed in Materials and Methods.

\subsubsection{Semi-classical treatment}
We now model a coherent receiver with the semi-classical treatment including the losses and coupler imperfections. The complex narrowband fields at the signal and LO ports are
\begin{align} \begin{aligned}
\begin{gathered}
    A_\mathrm{s}=C_\mathrm{s}e^{i(\omega_\mathrm{s}t+\phi_\mathrm{s})}\\
    A_\mathrm{LO}=C_\mathrm{LO}e^{i(\omega_\mathrm{LO}t+\phi_\mathrm{LO})}
\end{gathered}
\end{aligned} \end{align}
A lossless coupler with the unitary, U=$\frac{1}{\sqrt{2}}\left[\alpha\ \beta; -\beta\ \alpha\right]$, where $\alpha$, $\beta$ are complex numbers with $\abs{\alpha}^2+\abs{\beta}^2=2$ and $\mathrm{Im}{\{\alpha^*\beta\}}=0$, will do the following transformation.
\begin{align} \begin{aligned}
    \begin{bmatrix}
    A_{+}\\
    A_{-}
    \end{bmatrix}
    &=
    \frac{1}{\sqrt{2}}
    \begin{bmatrix}
    \alpha C_\mathrm{s}{ e^{{i(\omega_\mathrm{s} t+\phi_\mathrm{s})}}}+\beta C_\mathrm{LO}{ e^{{i(\omega_\mathrm{LO} t+\phi_\mathrm{LO}})}}\\
    -\beta C_\mathrm{s}{ e^{{i(\omega_\mathrm{s} t+\phi_\mathrm{s})}}}+\alpha C_\mathrm{LO}{ e^{{i(\omega_\mathrm{LO}} t+\phi_\mathrm{LO})})}
    \end{bmatrix}
\end{aligned} \end{align}
These fields will then be detected by a pair of photodiodes, generating a photocurrent proportional to power on each branch, $I_\mathrm{PD}=\left|E\right|^2R_\mathrm{PD}$, where $E=\mathrm{Re}\{A\}$. We normalize field amplitudes such that $\abs{A}^2$ gives optical power, and absorb the photodiode quantum efficiency $R_\mathrm{PD}$ and other scalar losses into an efficiency term $L$, giving $I_\mathrm{PD}=L\abs{A}^2$. To preserve the unitarity of the coupler transformation, we can also adjust $L$ accordingly.
Subtracting the photocurrents in the two branches gives
\begin{align} \begin{aligned}
    I_\mathrm{out}&=I_+-I_-=L\left[\abs{A_+}^2-\abs{A_-}^2\right]\\
    &=L\left[\Sigma (C_\mathrm{s}^2- C_\mathrm{LO}^2)+2\Pi C_\mathrm{s}C_\mathrm{LO}\cos(X(t))\right]
    \label{eq:PDcurrentu}
\end{aligned} \end{align}
where $\Sigma=\abs{\alpha}^2-\abs{\beta}^2$ and $\Pi=2\abs{\alpha}\abs{\beta}$, $X(t)=\omega_\mathrm{IF}t+\phi_\mathrm{s}-\phi_\mathrm{LO}$ and $\omega_\mathrm{IF}=\omega_\mathrm{s}-\omega_\mathrm{LO}$.

Common-mode rejection ratio ($\mathrm{CMRR}$) is defined as $\mathrm{CMRR}=\frac{P_\mathrm{unb}}{P_\mathrm{bal}}$ where $P_\mathrm{unb}$ is the electrical power when the photodiode currents are summed and $P_\mathrm{bal}$ is the electrical power when the photodiode currents are subtracted. Then, we can define
\begin{align} \begin{aligned}
    \label{eq:CMRR1}
    \mathrm{CMRR}=\frac{(\abs{A_+}^2+\abs{A_-}^2)^2}{(\abs{A_+}^2-\abs{A_-}^2)^2}=\frac{\Sigma^2+\Pi^2}{\Sigma^2}\approx \frac{\Pi^2}{\Sigma^2}
\end{aligned} \end{align}
where we approximated the last expression for high $\mathrm{CMRR}$. Substituting in this definition,
\begin{align} \begin{aligned}
    I_\mathrm{out}&=\Pi L\left[\frac{1}{\sqrt{\mathrm{CMRR}}} (C_\mathrm{s}^2- C_\mathrm{LO}^2)+2 C_\mathrm{s}C_\mathrm{LO}\cos(X(t))\right]
\end{aligned} \end{align}
For high $\mathrm{CMRR}$, we get a current proportional to the beat term.
\begin{align} \begin{aligned}
    I_\mathrm{out}&\approx2\Pi L C_\mathrm{s}C_\mathrm{LO}\cos(X(t))
\end{aligned} \end{align}
Now, we can replace $C$ with $S$ in \eqref{eq:S} and add the signal and noise carrying components in \eqref{eq:Ss}. There is also electronic noise, $\Delta i_\mathrm{n}$, coming from the electronic readout usually limited by the thermal noise in the amplifier.
\begin{align} \begin{aligned}
    I_\mathrm{out}&=\Pi L\Big[\frac{1}{\sqrt{\mathrm{CMRR}}} (s_\mathrm{s}+\Delta s_\mathrm{s}-s_\mathrm{LO}+\Delta s_\mathrm{LO})\\
    &+2 \sqrt{{(s_\mathrm{s}+\Delta s_\mathrm{s})(s_\mathrm{LO}}+\Delta s_\mathrm{LO})}\cos(X(t))\Big]+\Delta i_\mathrm{n}
\end{aligned} \end{align}
Since we are only interested in the signal and noise at $\omega_\mathrm{IF}$, we can filter the DC terms. We can also downconvert to DC for mean and variance calculations.
\begin{align} \begin{aligned}
    I_\mathrm{out}&=\Pi L\Big[\frac{1}{\sqrt{\mathrm{CMRR}}}(\Delta s_\mathrm{s}- \Delta s_\mathrm{LO})\\
    &+2 \sqrt{(s_\mathrm{s}+\Delta s_\mathrm{s})(s_\mathrm{LO}+\Delta s_\mathrm{LO})}\Big]+\Delta i_\mathrm{n}
\end{aligned} \end{align}
Then, the signal power is
\begin{align} \begin{aligned}
    P_\mathrm{sig}&=\langle I_\mathrm{out}\rangle^2\\
    &=4\Pi^2 L^2 \left\langle\sqrt{(s_\mathrm{s}+\Delta s_\mathrm{s})}\sqrt{(s_\mathrm{LO}+\Delta s_\mathrm{LO})}\right\rangle^2\\
    &=4\Pi^2 L^2\Bigg\langle \left(\sqrt{s_\mathrm{s}}+\frac{\Delta s_\mathrm{s}}{2\sqrt{s_\mathrm{s}}}-\frac{\Delta s_\mathrm{s}^2}{8s_\mathrm{s}^{3/2}}...\right)\\
    &\left(\sqrt{s_\mathrm{LO}}+\frac{\Delta s_\mathrm{LO}}{2\sqrt{s_\mathrm{LO}}}-\frac{\Delta s_\mathrm{LO}^2}{8s_\mathrm{LO}^{3/2}}...\right) \Bigg\rangle^2\\
    &\approx4\Pi^2 L^2s_\mathrm{s}s_\mathrm{LO}
\end{aligned} \end{align}
where we Taylor expanded the square root terms up to the second order, assumed high LO power limit, $s_\mathrm{LO}\gg s_\mathrm{s}$ and used the fact that $\langle \Delta s\rangle=0$. Now, the noise power is
\begin{align} \begin{aligned}
    P_\mathrm{noise}&=\langle \Delta I_\mathrm{out}^2\rangle=\langle I_\mathrm{out}^2\rangle-\langle I_\mathrm{out}\rangle^2\\
    &=\Pi^2 L^2\Bigg[\frac{1}{\mathrm{CMRR}}\left(\langle \Delta s_\mathrm{s}^2 \rangle+\langle\Delta s_\mathrm{LO}^2\rangle+2\langle\Delta s_\mathrm{s}\Delta s_\mathrm{LO}\rangle\right)\\
    &+4\left(\frac{s_\mathrm{s}\langle\Delta s_\mathrm{LO}^2\rangle}{4s_\mathrm{LO}}+\frac{s_\mathrm{LO}\langle\Delta s_\mathrm{s}^2\rangle}{4s_\mathrm{s}}+\frac{\langle\Delta s_\mathrm{s} \Delta s_\mathrm{LO}\rangle}{2}\right)\Bigg]\\
    &+\langle\Delta i_\mathrm{n}^2\rangle
\end{aligned} \end{align}
Taking the high LO power limit, $s_\mathrm{LO}\gg s_\mathrm{s}$.
\begin{align} \begin{aligned}
    \label{eq:Pn}
    P_\mathrm{noise}&= \Pi^2 L^2\left(\frac{1}{\mathrm{CMRR}}\langle\Delta s_\mathrm{LO}^2\rangle+\frac{s_\mathrm{LO}\langle\Delta s_\mathrm{s}^2\rangle}{s_\mathrm{s}}\right)+\langle\Delta i_\mathrm{n}^2\rangle
\end{aligned} \end{align}
Since LO is a coherent state, we replace $\langle\Delta s_\mathrm{LO}^2\rangle=\frac{\hbar\omega}{T} s_\mathrm{LO}$.
\begin{align} \begin{aligned}\label{eq:Pnoise}
    P_\mathrm{noise}&= \Pi^2 L^2\left(\frac{\hbar\omega}{T}\right) s_\mathrm{LO}\left(\frac{1}{\mathrm{CMRR}}+\frac{\langle\Delta s_\mathrm{s}^2\rangle T}{\hbar\omega s_\mathrm{s}}\right)+\langle\Delta i_\mathrm{n}^2\rangle
\end{aligned} \end{align}
We replace $\langle\Delta s_{s}^2\rangle=\frac{\hbar\omega}{T} s_{s}$, and it is also convenient in \eqref{eq:Pnoise} to normalize everything by LO power.
\begin{align} \begin{aligned}
\label{eq:Pnoisee}
    \frac{P_\mathrm{noise}}{s_\mathrm{LO}}&= \Pi^2 L^2\left(\frac{\hbar\omega}{T}\right) \left(\frac{1}{\mathrm{CMRR}}+1\right)+\frac{\langle\Delta i_\mathrm{n}^2\rangle}{s_\mathrm{LO}}
\end{aligned} \end{align}
For high $s_\mathrm{LO}$ and $\mathrm{CMRR}$, $\mathrm{SNR}$ for coherent detection is
\begin{align} \begin{aligned}
    \mathrm{SNR_{cd}}=\frac{P_\mathrm{sig}}{P_\mathrm{noise}}&\approx \frac{4s_\mathrm{s}^2}{\langle\Delta s_\mathrm{s}^2\rangle}=\frac{4s_\mathrm{s}}{\hbar\omega/T}
\end{aligned} \end{align}
We note that the semi-classical treatment uses the intensity noise $\langle\Delta s_\mathrm{s}^2\rangle$, which couples both quadratures of the signal field. This model is therefore valid only for coherent states. For non-classical states such as squeezed states, the quantum treatment is required.

\subsubsection{Quantum treatment}
We can do the same with the quantum treatment. Since \textbf{$\hat{E}$} can be defined in terms of $\hat{a}, \hat{a}^\dagger$, it suffices to study the transformation of a coherent receiver on $\hat{a}$, which will take the place of the $A$ from the semi-classical treatment. We define $\hat{a}_\mathrm{s}$ and $\hat{a}_\mathrm{LO}$ as the field operators for the signal and LO fields, respectively. These fields go through the coupler transformation, U=$\frac{1}{\sqrt{2}}\left[\alpha\ \beta; -\beta\ \alpha\right]$, where $\alpha$, $\beta$ are complex numbers, $\mathrm{Im}{\{\alpha^*\beta\}}=0$ and $\abs{\alpha}^2+\abs{\beta}^2=2$.
\begin{align} \begin{aligned}
    \begin{bmatrix}
    \hat{a}_{+}\\
    \hat{a}_{-}
    \end{bmatrix}
    &=
    \frac{1}{\sqrt{2}}
    \begin{bmatrix}
    \alpha \hat{a}_\mathrm{s}+\beta \hat{a}_\mathrm{LO}
    \\
    -\beta \hat{a}_\mathrm{s}+\alpha\hat{a}_\mathrm{LO}
    \end{bmatrix}
\end{aligned} \end{align}
Assuming ensemble measurements, we define a current operator, $\hat{I}=R\hat{S}$, that corresponds to the photocurrent from the photodetectors, using the previously defined power operator in \eqref{eq:power} multiplied by responsivity, $R$, which is a constant scalar. There is additional loss from the non-unity quantum efficiency of the photodiodes that can be modeled as interfering $\hat{a}_\pm$ with vacuum modes. Since this operation commutes with the initial coupler operation, we instead lump in these losses into $\hat{a}_\mathrm{s}, \hat{a}_\mathrm{LO}$ and into $\alpha, \beta$ in the coupler transformation. Then,
\begin{align} \begin{aligned}
    \hat{I}_\mathrm{out}&=\hat{S}_+-\hat{S}_-=R\frac{\hbar\omega}{T}(\hat{a}_+^\dagger\hat{a}_+-\hat{a}_-^\dagger\hat{a}_-)\\
    &=R\frac{\hbar\omega}{T}\left[\Sigma(\hat{a}_\mathrm{s}^{\dagger}\hat{a}_\mathrm{s}-\hat{a}_\mathrm{LO}^{\dagger}\hat{a}_\mathrm{LO})+\Pi\left( \hat{a}_\mathrm{s}^{\dagger}\hat{a}_\mathrm{LO}+\hat{a}_\mathrm{LO}^{\dagger}\hat{a}_\mathrm{s} \right)\right]
    \label{eq:Iqu}
\end{aligned} \end{align}
where $\Sigma=\abs{\alpha}^2-\abs{\beta}^2$ and $\Pi=2\abs{\alpha}\abs{\beta}$. Substituting in $\mathrm{CMRR}$ from \eqref{eq:CMRR1},
\begin{align} \begin{aligned}
    \hat{I}_\mathrm{out}&=\Pi R\frac{\hbar\omega}{T}\Bigg[\frac{1}{\sqrt{\mathrm{CMRR}}}(\hat{a}_\mathrm{s}^{\dagger}\hat{a}_\mathrm{s}-\hat{a}_\mathrm{LO}^{\dagger}\hat{a}_\mathrm{LO})\\
    &+\left( \hat{a}_\mathrm{s}^{\dagger}\hat{a}_\mathrm{LO}+\hat{a}_\mathrm{LO}^{\dagger}\hat{a}_\mathrm{s} \right)\Bigg]
\end{aligned} \end{align}
We can now add the signal and noise-carrying components in \eqref{eq:nsq}. We also add an operator, $\Delta \hat{i}_\mathrm{n}=\Delta i_\mathrm{n} . \hat{I}$, which is a product of a random scalar with an expected value of zero representing the electronic noise, $\Delta i_\mathrm{n}$ and the identity operator, $\hat{I}$.
\begin{align} \begin{aligned}
    \hat{I}_\mathrm{out}&=\Pi R\frac{\hbar\omega}{T}\Bigg[\frac{1}{\sqrt{\mathrm{CMRR}}}(\hat{a}_\mathrm{s}^{\dagger}\hat{a}_\mathrm{s}-\hat{a}_\mathrm{LO}^{\dagger}\hat{a}_\mathrm{LO})\\
    &+\left( \hat{a}_\mathrm{s}^{\dagger}\hat{a}_\mathrm{LO}+\hat{a}_\mathrm{LO}^{\dagger}\hat{a}_\mathrm{s} \right)\Bigg]+\Delta \hat{i}_\mathrm{n}\\
    &=\Pi R\frac{\hbar\omega}{T}\Bigg[\frac{1}{\sqrt{\mathrm{CMRR}}}\Big(q_\mathrm{s}^2+2q_\mathrm{s}\Delta \hat{q}_\mathrm{s}+\Delta \hat{p}_\mathrm{s}^2-q_\mathrm{LO}^2\\
    &-2q_\mathrm{LO}\Delta \hat{q}_\mathrm{LO}-\Delta \hat{p}_\mathrm{LO}^2\Big)+2\Big(q_\mathrm{s}q_\mathrm{LO}+q_\mathrm{s}\Delta \hat{q}_\mathrm{LO}\\
    &+q_\mathrm{LO}\Delta\hat{q}_\mathrm{s}+\Delta\hat{q}_\mathrm{s}\Delta \hat{q}_\mathrm{LO}+\Delta \hat{p}_\mathrm{s}\Delta \hat{p}_\mathrm{LO}\Big)\Bigg]+\Delta \hat{i}_\mathrm{n}
    \label{eq:Iqout}
\end{aligned} \end{align}
where we used the fact that $\hat{q}, \hat{p}$ are Hermitian. Assuming coherent states for LO and signal, assuming high LO power limit, $q_\mathrm{LO}\gg q_\mathrm{s}$, and assuming high $\mathrm{CMRR}$
\begin{align} \begin{aligned}
    \hat{I}_\mathrm{out}&\approx2\Pi R\frac{\hbar\omega}{T} q_\mathrm{LO}\left(q_\mathrm{s}+\Delta\hat{q}_\mathrm{s}\right)=2\Pi R\frac{\hbar\omega}{T} q_\mathrm{LO}\hat{q}_\mathrm{s}
\end{aligned} \end{align}
gives the quadrature operator of the signal field, meaning the photocurrent follows the quadrature statistics. The signal power can then be defined as
\begin{align} \begin{aligned}
    P_\mathrm{sig}&=I_\mathrm{out}^2=4\Pi^2R^2\left(\frac{\hbar\omega}{T}\right)^2 q_\mathrm{LO}^2q_\mathrm{s}^2\approx4\Pi^2R^2s_\mathrm{LO}s_\mathrm{s}
\end{aligned} \end{align}
Without any assumptions, going back to \eqref{eq:Iqout}, the noise power is
\begin{align} \begin{aligned}
    P_\mathrm{noise}&=\langle \Delta \hat{I}_\mathrm{out}^2\rangle=
    \Pi^2R^2\left(\frac{\hbar\omega}{T}\right)^2\Bigg[\frac{1}{\mathrm{CMRR}}\Big(4q_\mathrm{s}^2\langle\Delta \hat{q}_\mathrm{s}^2\rangle\\
    &+4q_\mathrm{LO}^2\langle\Delta \hat{q}_\mathrm{LO}^2\rangle+Q_\mathrm{u}\Big)\\
    &+4q_\mathrm{s}^2\langle\Delta \hat{q}_\mathrm{LO}^2\rangle+4q_\mathrm{LO}^2\langle\Delta \hat{q}_\mathrm{s}^2\rangle+Q_\mathrm{b}\Bigg]+\langle\Delta \hat{i}_\mathrm{n}^2\rangle
\end{aligned} \end{align}
where $Q_\mathrm{u}=-2\langle\Delta \hat{q}_\mathrm{s}^2\rangle\langle\Delta \hat{q}_\mathrm{LO}^2\rangle+\langle\Delta \hat{p}_\mathrm{LO}^4\rangle+\langle\Delta \hat{p}_\mathrm{s}^4\rangle+8q_\mathrm{s}q_\mathrm{LO}(\langle\Delta \hat{p}_\mathrm{s}^2\rangle-\langle\Delta \hat{p}_\mathrm{LO}^2\rangle)$ and $Q_\mathrm{b}=4\langle\Delta \hat{q}_\mathrm{s}^2\rangle\langle\Delta \hat{q}_\mathrm{LO}^2\rangle-4\langle\Delta \hat{p}_\mathrm{s}^2\rangle\langle\Delta \hat{p}_\mathrm{LO}^2\rangle$ are the higher order terms assuming Gaussian states for LO and signal. Neglecting the higher order terms,
\begin{align} \begin{aligned}
    P_\mathrm{noise}&\approx4\Pi^2R^2\left(\frac{\hbar\omega}{T}\right)^2\Bigg[\frac{1}{\mathrm{CMRR}}\left(q_\mathrm{s}^2\langle\Delta \hat{q}_\mathrm{s}^2\rangle+q_\mathrm{LO}^2\langle\Delta \hat{q}_\mathrm{LO}^2\rangle\right)\\
    &+q_\mathrm{s}^2\langle\Delta \hat{q}_\mathrm{LO}^2\rangle+q_\mathrm{LO}^2\langle\Delta \hat{q}_\mathrm{s}^2\rangle\Bigg]+\langle\Delta \hat{i}_\mathrm{n}^2\rangle
\end{aligned} \end{align}
Taking the high LO power limit, $q_\mathrm{LO}\gg q_\mathrm{s}$,
\begin{align} \begin{aligned}\label{eq:Pnoise2}
    P_\mathrm{noise}&\approx4\Pi^2R^2\left(\frac{\hbar\omega}{T}\right)^2\\
    &\Bigg(\frac{1}{\mathrm{CMRR}}q_\mathrm{LO}^2\langle\Delta \hat{q}_\mathrm{LO}^2\rangle+q_\mathrm{LO}^2\langle\Delta \hat{q}_\mathrm{s}^2\rangle\Bigg)+\langle\Delta \hat{i}_\mathrm{n}^2\rangle
\end{aligned} \end{align}
Normalizing by LO power $s_\mathrm{LO}\approx \left(\frac{\hbar\omega}{T}\right)q_\mathrm{LO}^2$ in \eqref{eq:Pnoise2},
\begin{align} \begin{aligned}
    \frac{P_\mathrm{noise}}{s_\mathrm{LO}}&\approx4\Pi^2R^2\left(\frac{\hbar\omega}{T}\right)\\
    &\Bigg(\frac{1}{\mathrm{CMRR}}\langle\Delta \hat{q}_\mathrm{LO}^2\rangle+\langle\Delta \hat{q}_\mathrm{s}^2\rangle\Bigg)+\frac{\langle\Delta \hat{i}_\mathrm{n}^2\rangle}{s_\mathrm{LO}}
\end{aligned} \end{align}
Assuming coherent states for signal and LO,  $\langle(\Delta \hat{q})^2\rangle=\langle(\Delta \hat{p})^2\rangle=\frac{1}{4}$, we arrive at the same expression as the semi-classical treatment in \eqref{eq:Pnoisee}, showing the equivalence of semi-classical and quantum treatments when coherent states are assumed.
\begin{align} \begin{aligned}\label{eq:Pnoisee2}
    \frac{P_\mathrm{noise}}{s_\mathrm{LO}}&\approx\Pi^2R^2\left(\frac{\hbar\omega}{T}\right)\Bigg(\frac{1}{\mathrm{CMRR}}+1\Bigg)+\frac{\langle\Delta \hat{i}_\mathrm{n}^2\rangle}{s_\mathrm{LO}}
\end{aligned} \end{align}
For high $s_\mathrm{LO}$ and $\mathrm{CMRR}$, $\mathrm{SNR}$ for coherent detection is
\begin{align} \begin{aligned}
    \mathrm{SNR_{cd}}=\frac{\mathrm{P_{sig}}}{\mathrm{P_{noise}}}\approx \frac{q_\mathrm{s}^2}{\langle\Delta \hat{q}_\mathrm{s}^2\rangle}
\end{aligned}=4q_\mathrm{s}^2=\frac{4s_\mathrm{s}}{\hbar\omega/T} \end{align}

The equivalence of semi-classical and quantum treatments for coherent states shows that the semi-classical model is good enough for communications with classical states, but the complete quantum treatment with the annihilation-creation operator is required when discussing communications with non-classical states.

\subsection{Quantum-limited coherent receiver design}
\begin{figure*}
    \centering
    \includegraphics[width=\linewidth]{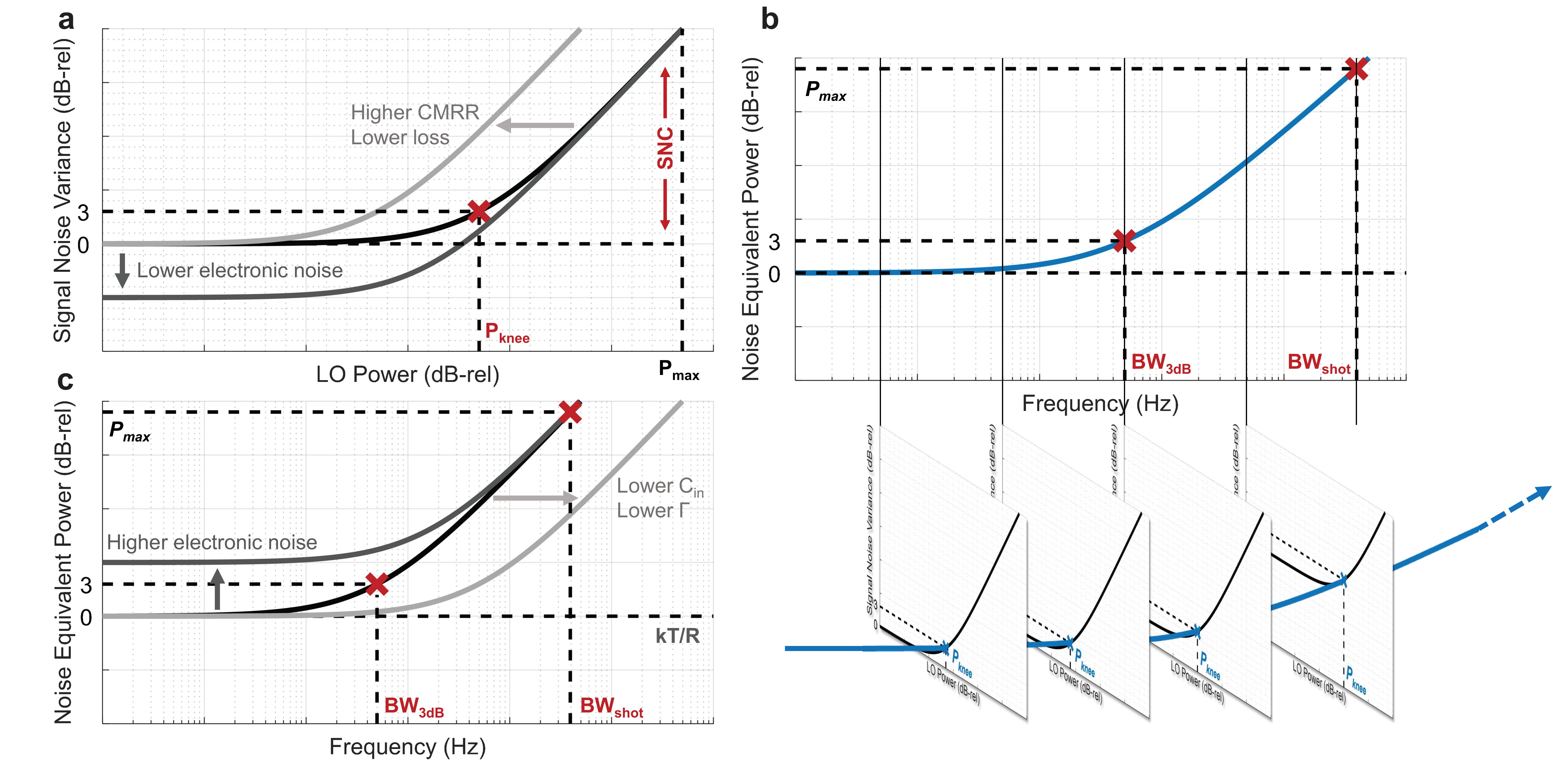}
    \caption{\textbf{Design guide for quantum coherent receivers.} \textbf{a)} Normalized signal noise variance in \unit{dB} as a function of LO power, defined as the normalized total electronic noise power at the output integrated over a given bandwidth. $P_{\text{knee}}$ is the LO power at which signal noise variance reaches 3\,\unit{dB} above the electronic noise floor, marking the transition from electronic-noise-limited to shot-noise-limited operation. The shot noise clearance ($\mathrm{SNC}$) is the ratio between the noise power at the maximum LO power ($P_\mathrm{max}$) and the electronic noise floor. Higher $\mathrm{CMRR}$ and lower loss shift the curve left (light gray) leading to a higher $\mathrm{SNC}$ and lower electronic noise shifts it down (dark gray). \textbf{b)} Normalized noise equivalent power (NEP) in \unit{dB} as a function of frequency. NEP is the input optical power required for an output $\mathrm{SNR}$ of 1. NEP rises with frequency until at $\mathrm{BW_{shot}}$ where the required power reaches $P_\mathrm{max}$. Beyond this, the system is no longer shot noise limited. $\mathrm{BW_{3 dB}}$ marks where NEP has risen 3\,\unit{dB} above its low-frequency value. Another way to see this is plotting $P_\mathrm{knee}$ values across all frequencies as shown below. \textbf{c)} NEP as a function of frequency as in (b), additionally showing design parameter dependence. Lower input capacitance ($C_{\text{in}}$) and transistor channel noise ($\Gamma$) shift the curve right (light gray), extending $\mathrm{BW_{3 dB}}$ and $\mathrm{BW_{shot}}$, higher electronic noise shifts it up (dark gray).}
    \label{fig:fig2}
\end{figure*}

With the coherent receiver models, we can define a few key specifications that characterize the performance of a quantum-limited coherent receiver, illustrated in Fig.~\ref{fig:fig2}. These specifications are the insertion loss, common-mode rejection ratio ($\mathrm{CMRR}$), shot noise clearance ($\mathrm{SNC}$), LO power knee ($\mathrm{P_{knee}}$), 3-\unit{dB} bandwidth ($\mathrm{BW_{3dB}}$) and shot-noise-limited bandwidth ($\mathrm{BW_{shot}}$) \cite{Gurses2023, Gurses2022, Bruynsteen2021, Tasker2021, Tasker2024}. We will use the semi-classical model to define these variables since it is simpler to connect to measurements. However, as outlined above, both models are equivalent.

Using \eqref{eq:Pn}, we can write the noise power again, but now, we also consider a noisy LO above the shot noise limit, setting $\langle(\Delta s_\mathrm{LO})^2\rangle=s_\mathrm{LO}^2\mathrm{RIN}$, where RIN is the relative intensity noise spectral density of the LO that converges to $\mathrm{RIN}=\frac{\hbar\omega}{s_\mathrm{LO}T}$ for a shot-noise-limited LO. Then,
\begin{align} \begin{aligned}
    P_\mathrm{noise}&= \Pi^2 L^2\frac{\hbar\omega}{T} s_\mathrm{LO}\left(1+\frac{s_\mathrm{LO}\mathrm{RIN}}{\hbar\omega \mathrm{CMRR}}\right)+\langle(\Delta i_\mathrm{n})^2\rangle
\end{aligned} \end{align}
In this expression, the first term signifies noise contribution from the signal shot noise, the second term signifies noise contribution from the noisy LO, and the third term signifies noise contribution from electronics. For notational convenience in the proportionality expressions that follow, we write $i_\mathrm{n}^2\equiv\langle(\Delta i_\mathrm{n})^2\rangle$ for the input-referred electronic noise variance. Dividing $P_\mathrm{noise}$ by $P_\mathrm{LO}$ yields a convenient expression:
\begin{align} \begin{aligned}
\label{eq:iq}
    \frac{P_\mathrm{noise}}{P_\mathrm{LO}}\propto L^2\left(1+\frac{\mathrm{RIN}\cdot P_\mathrm{LO}}{\mathrm{CMRR}}\right)+\frac{i_\mathrm{n}^2}{P_\mathrm{LO}}
\end{aligned} \end{align}

At high $\mathrm{CMRR}$, noise power starts at the electronic noise floor and increases linearly with LO power above a certain LO power. We define this LO power as the LO power knee, $P_\mathrm{knee}$. More precisely, $P_\mathrm{knee}$ is the power at which the noise contribution of the signal shot noise is equal to other noise contributions, i.e., the LO power required to have 3~\unit{dB} shot noise clearance (at which the QRX is shot noise limited). For a shot-noise-limited LO, $P_\mathrm{knee}$ can be defined as
\begin{align}
    P_\mathrm{knee}=\frac{i_\mathrm{n}^2}{\Pi^2L^2\frac{\hbar\omega}{T}\left(1-\frac{1}{\mathrm{CMRR}}\right)}\propto\frac{i_\mathrm{n}^2}{L^2}
\end{align}
We note that, at low $\mathrm{CMRR}$, $P_\mathrm{knee}$ is high since the LO shot noise starts to scale proportionally to the signal shot noise, introducing decoherence to the downconverted signal \cite{Gurses2022}. At high $\mathrm{CMRR}$, above $P_\mathrm{knee}$, the receiver noise floor is limited by the shot noise of the signal field only, eliminating the noise contribution from the noisy LO, and allowing quantum coherent receivers to probe quadrature statistics of the signal field. The $P_\mathrm{knee}$ should be minimized to reduce the QRX power consumption and ensure enough LO power can be supplied to the system as more QRXs are integrated on-chip.

At the maximum LO power, $P_\mathrm{max}$, the receiver usually saturates due to nonlinear effects, such as two-photon absorption in the waveguides or the carrier screening effect in photodiodes. The shot noise clearance is defined as the ratio of the total noise power from \eqref{eq:iq} at $P_\mathrm{max}$ to the noise power at no LO power, assuming the QRX is operated in the signal shot-noise-limited regime with a high enough $\mathrm{CMRR}$ \cite{Gurses2022}:
\begin{align}
\label{eq:SNC}
    \mathrm{\mathrm{SNC}}\propto\frac{L^2P_\mathrm{max}}{i_\mathrm{n}^2}+1\approx\frac{P_\mathrm{max}}{{P_\mathrm{knee}}}+1
\end{align}
Maximizing the shot noise clearance is crucial to probe the quadrature statistics of the signal field accurately.

We can also write the noise power as a function of frequency, $\omega_\mathrm{IF}$. In this definition, $\mathrm{CMRR}$, including $\Sigma$, $\Pi$, and electronic noise, should also be written as a function of $\omega_\mathrm{IF}$. Since within the electronic bandwidth of the receiver, the optical frequency stays approximately constant, we assume $\hbar\omega$ to remain constant.
\begin{align} \begin{aligned}
    P_\mathrm{noise}(\omega_\mathrm{IF})&= [\Pi(\omega_\mathrm{IF})]^2 L^2\frac{\hbar\omega}{T} s_\mathrm{LO}\left(\frac{1}{\mathrm{CMRR}(\omega_\mathrm{IF})}+1\right)\\
    &+\langle[\Delta i_\mathrm{n}(\omega_\mathrm{IF})]^2\rangle
\end{aligned} \end{align}
Due to the frequency-dependent response of the QRX, $L$ can be defined as a function of frequency, $L(f)$. The 3-\unit{dB} bandwidth sets the classical operation bandwidth of a QRX. The frequency at which loss increases by 3~\unit{dB} is defined as the 3-\unit{dB} cut-off frequency, namely $L(f_\mathrm{3dB})=L(0)/2$. At the 3-\unit{dB} cut-off frequency, the QRX output signal will be attenuated by 3~\unit{dB}. However, in the infinite $\mathrm{SNC}$ limit, this 3-\unit{dB} attenuation would not introduce decoherence to the quantum state and, for classical operation, does not change the $\mathrm{SNR}$. Shot-noise-limited bandwidth is the bandwidth at which the $\mathrm{SNC}$ is reduced to 3~\unit{dB} and at which the QRX stops being shot-noise-limited. For a frequency-dependent $L$, the shot-noise-limited cut-off frequency is defined as the frequency at which $P_\mathrm{knee}=P_{\mathrm{LO},\mathrm{max}}$, namely $L(f_\mathrm{shot})=i_\mathrm{n}^2/P_{\mathrm{LO},\mathrm{max}}$. Above this frequency, significant decoherence is introduced to the signal. Therefore, this is the bandwidth used for the quantum operation of the QRX.

The detection efficiency of the QRX determines how faithfully the receiver reproduces the quadrature statistics of the input field. We decompose the total detection efficiency into two contributions:
\begin{align}
    \eta = \eta_{\mathrm{opt}} \cdot \eta_{\mathrm{SNC}}
\end{align}
where $\eta_{\mathrm{opt}}$ is the optical efficiency, accounting for edge coupler loss, photodiode quantum efficiency, and routing loss, and $\eta_{\mathrm{SNC}}$ captures the finite shot noise clearance set by the LO power. At finite $\mathrm{SNC}$, the electronic noise floor contributes a vacuum-equivalent noise term that degrades the effective detection efficiency. From \eqref{eq:SNC}, this contribution is
\begin{align}
    \eta_{\mathrm{SNC}} = 1 - \frac{1}{\mathrm{SNC}} = 1 - \frac{P_{\mathrm{knee}}}{s_\mathrm{LO}}
\end{align}
giving
\begin{align}
\label{eq:eta}
    \eta = \eta_{\mathrm{opt}}\left(1 - \frac{P_{\mathrm{knee}}}{s_\mathrm{LO}}\right)
\end{align}
At infinite $\mathrm{SNC}$ ($s_\mathrm{LO} \gg P_{\mathrm{knee}}$), $\eta \to \eta_{\mathrm{opt}}$, and the detection efficiency is limited only by optical losses. At finite $\mathrm{SNC}$, increasing $s_\mathrm{LO}$ improves $\eta$ but adds to the total power consumption of the link.

For a squeezed state with squeezing parameter $r$, the measured quadrature variance including detection efficiency is
\begin{align}
\label{eq:Veff}
    \langle(\Delta \hat{q})^2\rangle_{\mathrm{meas}} = \frac{1}{4}\left(\eta e^{-2r} + 1 - \eta\right)
\end{align}
where the first term is the attenuated squeezed variance and the second is the vacuum noise mixed in by loss. This expression shows that both optical loss and finite $\mathrm{SNC}$ degrade the observable squeezing, and that the LO power directly controls the latter contribution through \eqref{eq:eta}.

\subsection{Integrated photonic-electronic quantum coherent receiver and large-scale integration}
\begin{figure}
    \centering
    \includegraphics[width=0.95\linewidth]{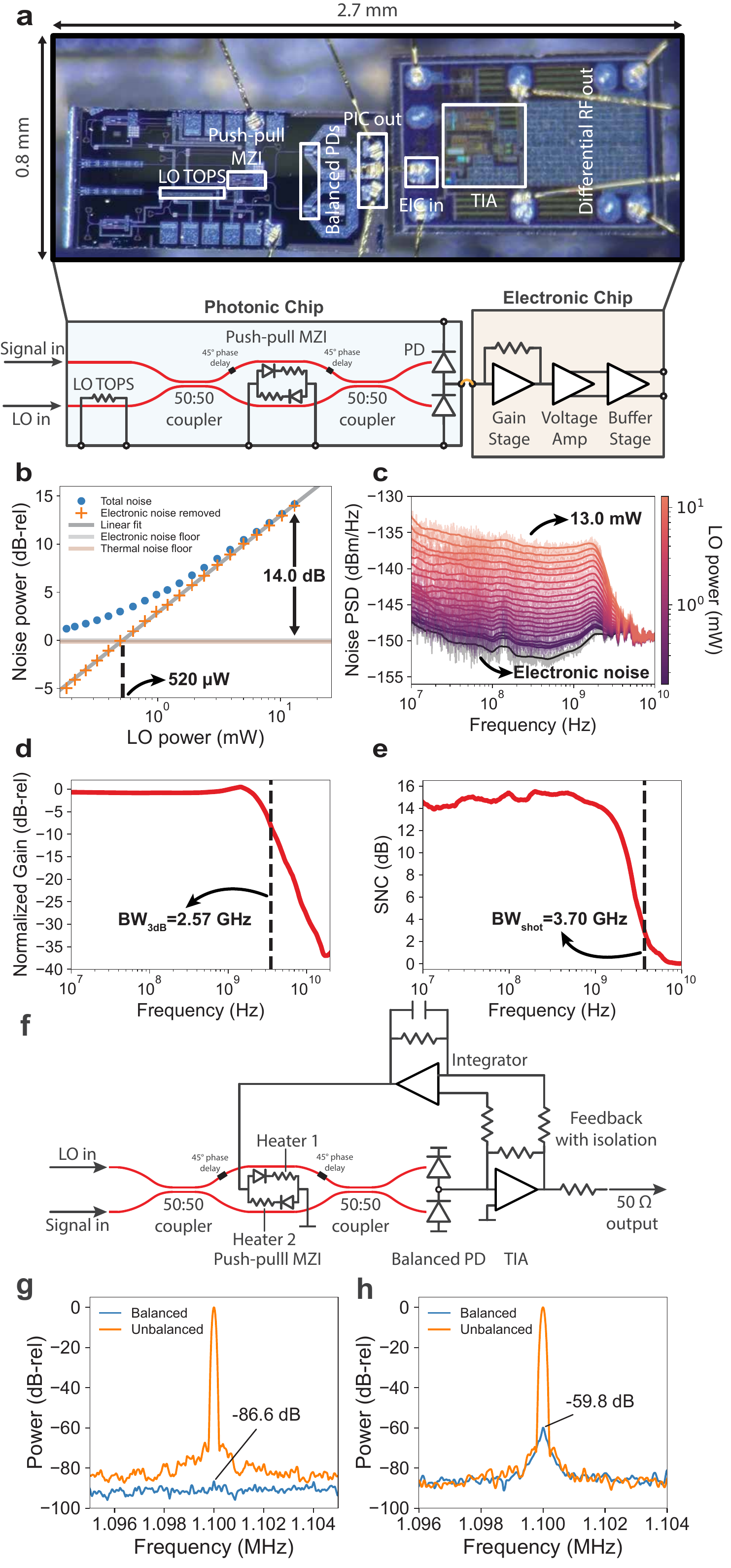}
    \caption{\textbf{Integrated photonic-electronic quantum coherent receiver. \textbf{a}} Micrograph of the PIC wirebonded to the EIC, with circuit schematic. \textbf{b} LO power sweep, showing a shot-noise-limited regime above the LO power knee ($P_\mathrm{knee}$) of 520 $\mathrm{\mu}$W with a shot noise clearance ($\mathrm{SNC}$) of 14 \unit{dB}. \textbf{c} Noise power spectral densities (PSDs) measured at different LO powers. \textbf{d} Frequency response showing a 3-\unit{dB} bandwidth of 2.57 GHz. \textbf{e} $\mathrm{SNC}$ frequency spectrum showing shot-noise limited bandwidth of 3.50 GHz. \textbf{f} $\mathrm{CMRR}$ correction scheme, where a fraction of the amplified photocurrent is fed back through an integrator to trim the MZI. \textbf{g} $\mathrm{CMRR}$ characterization, where the output measured with the MZI tuned to route all coupled light to one photodiode (unbalanced) vs. equal illumination of both photodiodes (balanced). \textbf{h} Measurement repeated at higher LO power for the balanced case, yielding a $\mathrm{CMRR}$ of 90.2 \unit{dB}.}
    \label{fig:fig3}
\end{figure}

With this design guide in place, we design a single photonic-electronic quantum coherent receiver for single channel high-bandwidth squeezed light detection, required for squeezed light communications. We then demonstrate a 32-channel array, by leveraging large-scale integration in photonic integrated circuits (PICs), and characterize all channels to demonstrate quantum-limited performance at scale.

The design and fabrication details of the PIC and EIC, along with the packaging, are described in Materials and Methods.

Bandwidth, $\mathrm{CMRR}$, and shot noise clearance measurement procedures are detailed in Materials and Methods. 

As seen in Fig.~\ref{fig:fig3}b and ~\ref{fig:fig3}c, the shot noise clearance is 14.0 \unit{dB} over the bandwidth of the QRX and the LO power knee ($P_\mathrm{knee}$) is 520 $\mathrm{\mu}$W. The $\mathrm{SNC}$ frequency response with maximum LO photocurrent is seen in Fig.~\ref{fig:fig3}e, showing a shot-noise-limited bandwidth of 3.50 GHz. A shot noise line is also fitted to the electronic noise subtracted measurements showing a near-unity slope of 1.007 $\mathrm{\pm}$ 0.015, confirming the shot-noise-limited noise floor. The optoelectronic response of the QRX is shown in Fig.~\ref{fig:fig3}d, showing a 3-\unit{dB} bandwidth of 2.57 GHz.

\begin{figure*}[t!]
    \centering
    \includegraphics[width=\linewidth]{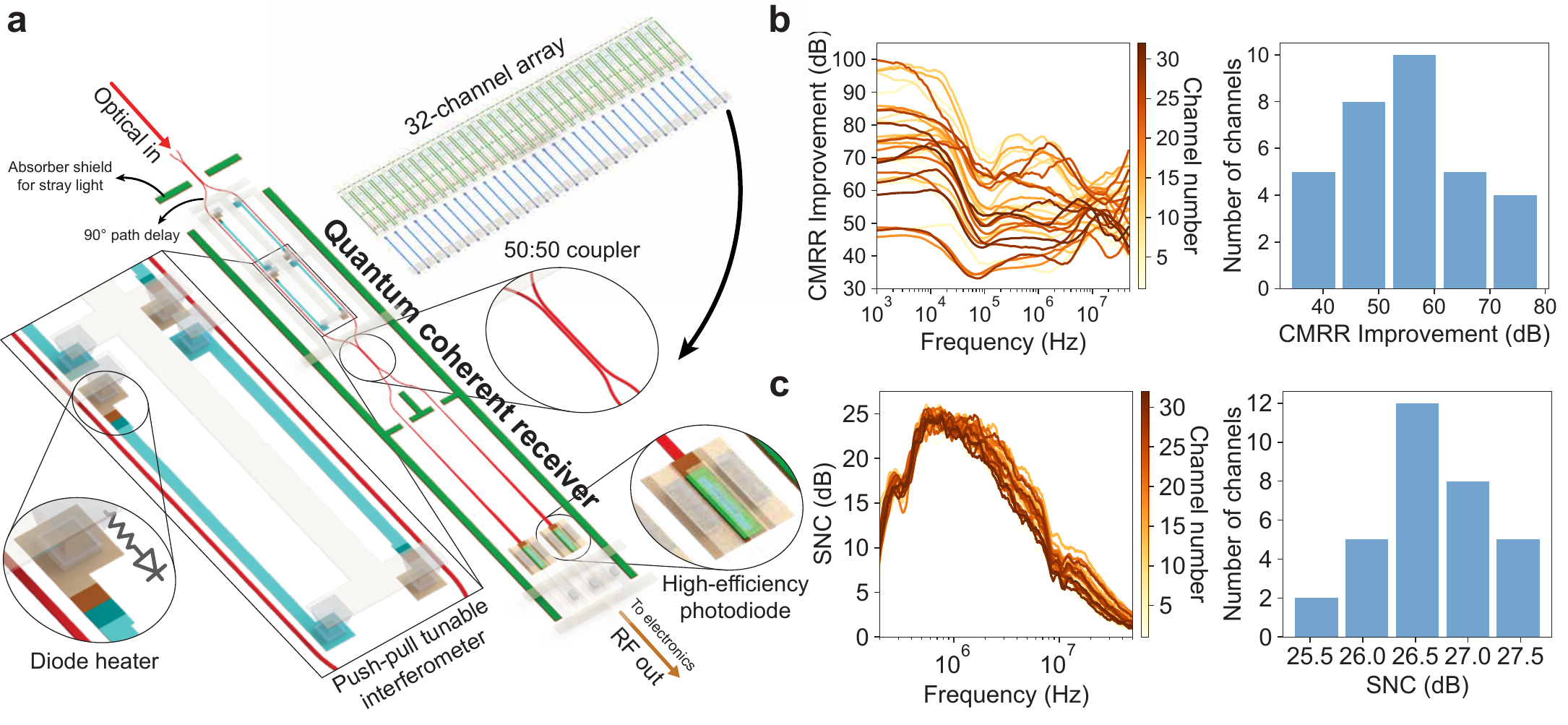}
    \caption{\textbf{Integrated quantum coherent receiver array} \textbf{a)} 32-channel quantum coherent receiver array design, showing the layout and different components. The optical input to the array is split across 32 channels, and the photocurrents from the outputs are separately sent to a 32-channel transimpedance amplifier (TIA) array along with 32 feedback loops for automatic $\mathrm{CMRR}$ correction at scale. \textbf{b)} 32-channel $\mathrm{CMRR}$ measurement. (Left) $\mathrm{CMRR}$ improvement in \unit{dB} from the automatic $\mathrm{CMRR}$ correction as a function of frequency for each individual channel. (Right) Histogram of CMMR improvement, which is the minimum $\mathrm{CMRR}$ for each channel, showing a median of 54.3 dB, a minimum of 33.7 dB, and a maximum of 79.1 dB. \textbf{c)} 32-channel shot noise clearance ($\mathrm{SNC}$) measurement. (Left) $\mathrm{SNC}$ in \unit{dB} as a function of frequency for each individual channel. (Right) Histogram of $\mathrm{SNC}$, showing a median of 26.6 dB, a minimum of 25.3 dB, and a maximum of 27.7 dB.}
    \label{fig:fig4}
\end{figure*}

Using the on-chip MZI to actively trim the MZI splitting ratio allows us to demonstrate automatic correction of the $\mathrm{CMRR}$. This is important to scale the receiver to many channels as fabrication variations across channels such as the coupling region gap variations of the 50:50 couplers and surface roughness can have a significant detriment to the uniformity of the $\mathrm{CMRR}$ across channels \cite{gurses2022large}. To that end, with an off-chip feedback loop, we demonstrate automatic correction of the $\mathrm{CMRR}$ with the circuit schematic shown in Fig.~\ref{fig:fig3}f.

The $\mathrm{CMRR}$ measurement procedure is described in Materials and Methods. The average $\mathrm{CMRR}$ from 10 traces measured over 10 seconds is 90.2 \unit{dB}, with a maximum $\mathrm{CMRR}$ of 92.3 dB.

Finally, we scale the PIC to 32 channels to demonstrate coherent and scalable quantum coherent receiver arrays, whose design is shown in Fig. \ref{fig:fig4}a. The 32-channel $\mathrm{CMRR}$ and $\mathrm{SNC}$ measurement procedures are described in Materials and Methods. As shown in Fig.~\ref{fig:fig4}b, the array exhibits strong and relatively uniform suppression, with a median $\mathrm{CMRR}$ improvement of 76.8 \unit{dB}, a minimum of 52.4 \unit{dB}, and a maximum of 104.2 \unit{dB}. All channels show a reduction in improvement between 10 kHz and 100 kHz, which is most likely set by the bandwidth of the integrator op-amp in the feedback path. The resulting $\mathrm{SNC}$ responses, shown in Fig.~\ref{fig:fig4}c, confirm that all channels operate well within the signal shot-noise-limited regime, with $\mathrm{SNC}$ exceeding 10 \unit{dB} throughout the array. The peak $\mathrm{SNC}$ values are tightly distributed, with a median of 26.6 \unit{dB}, a minimum of 25.3 \unit{dB}, and a maximum of 27.7 \unit{dB}. This uniform high-$\mathrm{SNC}$ operation indicates that the QRX architecture can be parallelized to very large channel counts without LO-power bottlenecks, in principle allowing thousands of channels to share a single LO input before reaching the two-photon-absorption power ceiling in silicon waveguides \cite{Chung2018}. The slightly lower median $\mathrm{SNC}$ relative to the single-channel receiver is attributed to channel-to-channel variation in LO power, electronic noise, and frequency response, as well as the lower allowable per-channel LO power set by early saturation in one array element \cite{Gurses2025OnChipPhasedArray,Gurses2024FreeSpaceQIP,Gurses2024CoherentImager}.

\subsection{Squeezed light measurements}

\begin{figure*}[t!]
    \centering
    \includegraphics[width=\linewidth]{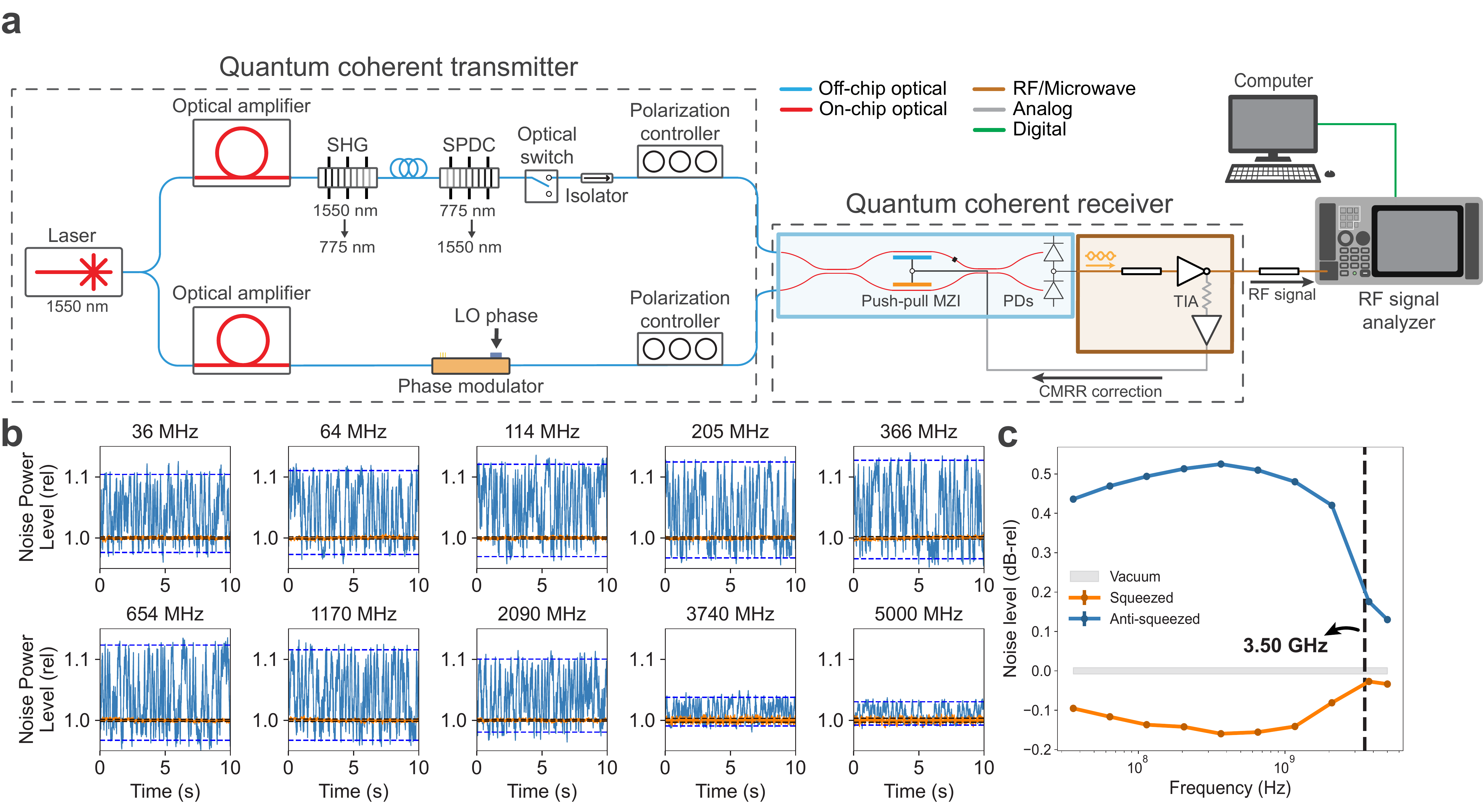}
    \caption{\textbf{Squeezed light measurements.} \textbf{a)} Experiment setup. A 1550 nm continuous-wave laser is divided by a 50-50 splitter into local oscillator (LO, bottom) and squeezer (signal, top) paths. The two paths are interfered at the on-chip tunable MZI and then fed to a balanced pair of Germanium photodiodes. The output current is fed to an on-chip amplifier. A portion of the output voltage signal is fed back to the MZIs for $\mathrm{CMRR}$ correction, while the remainder is routed to an RF signal analyzer. \textbf{b)} Time-domain traces. Output noise power is plotted vs. time as the LO phase is ramped at 1 Hz. The different subplots indicate noise integrated over a fixed bandwidth centered at each labeled frequency. \textbf{c)} Squeezed light frequency spectrum. Squeezed and anti-squeezed noise levels vs. frequency. The data is referenced to the vacuum noise level, obtained by opening the optical switch in the signal path.}
    \label{fig:fig5}
\end{figure*}

To demonstrate squeezed light detection across a high bandwidth, a transceiver was assembled in combination with the integrated QRX and a fiber-optic QTX as seen in Fig. \ref{fig:fig5}a. The experimental setup and QTX components are described in Materials and Methods.

Squeezed vacuum is injected into the QRX as signal and coherent light is injected as LO with a continuous-wave power of 9.50 mW, corresponding to an $\mathrm{SNC}$ of 12.7 \unit{dB} seen in Fig.~\ref{fig:fig3}b. Detailed measurement parameters are given in Materials and Methods.

10-second snippets of the data measured at different frequencies ranging from 36 MHz to 5 GHz are shown in Fig.~\ref{fig:fig5}b. The determined noise levels for squeezed and anti-squeezed quadratures normalized to the shot noise level (SNL) with their respective error bars at each frequency are shown in Fig.~\ref{fig:fig5}b. The noise level estimation procedure is described in Materials and Methods. Across the entire dataset, a maximum squeezing level of $0.15\pm0.01$ \unit{dB} below the SNL and a maximum antisqueezing level of $0.52\pm0.01$ \unit{dB} above the SNL was measured at a frequency of 366 MHz. From this estimate, we obtain $r = 0.661^{-0.041}_{+0.043} $ and $\eta = 0.046^{+0.005}_{-0.004}$. This corresponds to a system loss of 13.3 \unit{dB}.

While we were limited by the source and tabletop component losses in the squeezed light measurements, on-chip loss sets the bound on how much squeezing can be observed with the QRX. The on-chip system loss comprises the optical losses and the optoelectronic loss determined by the shot noise clearance and PD quantum efficiency (QE). The QRX has a total optical loss of 2.7 \unit{dB} with 1.3 \unit{dB} from edge couplers, 1.4 \unit{dB} from PD QE, and a negligible amount of loss from the TOPS, MZI, and routing. As shown in Fig.~\ref{fig:fig3}, the shot noise clearance is also greater than 10 \unit{dB} up to 2.24 GHz. Therefore, the system loss is at most 3 \unit{dB} over the bandwidth of the receiver, enabling sensitivities of 3 \unit{dB} below the SNL.

\subsection{Communications with squeezed light}

We now formally introduce the Shannon and Holevo limits as bounds on the channel capacity and explore how to design links that can approach the Holevo limit. The complex amplitude of a bosonic mode is $\hat{a} = \hat{q} + i\hat{p}$ \eqref{eq:a_q_p}, where $\hat{q}$ and $\hat{p}$ are the two conjugate quadratures and $\langle N\rangle = \langle \hat{a}^\dagger\hat{a}\rangle$ is the mean photon number of the mode. The Shannon capacity limit depends on whether one or both quadratures are used for communication \cite{Banaszek2020}. When information is encoded in a single quadrature and decoded with coherent detection, the one-quadrature Shannon capacity per mode is
\begin{align}
\label{eq:CS1}
    C_{\mathrm{S1}} = \frac{1}{2}\log_2\left(1 + 4\langle N\rangle\right)
\end{align}
When both quadratures are used with heterodyne detection, the two-quadrature Shannon capacity per mode is
\begin{align}
    C_{\mathrm{S2}} = \log_2(1 + \langle N\rangle)
    \label{eq:Shannon}
\end{align}
Allowing arbitrary quantum measurements, including joint detection over many channel uses, raises the bound on the accessible classical information per mode to the Holevo limit
\begin{align}
    C_{\mathrm{Hol}} = g(\langle N\rangle) = (\langle N\rangle+1)\log_2(\langle N\rangle+1) - \langle N\rangle\log_2\langle N\rangle
    \label{eq:Holevo}
\end{align}
where $g(\cdot)$ is the bosonic entropy function \cite{Holevo1973}. The Holevo limit strictly exceeds both Shannon limits for all $\langle N\rangle > 0$ \cite{GursesInterconnects2026}.

The gap between the Shannon and Holevo limits arises from superadditive coding gain. For a codeword of $n$ transmitted symbols, classical symbol-by-symbol detection measures each received mode independently and the total information extracted is the sum over all symbols. Since coherent states are not orthogonal, nearby constellation points always have some quantum mechanical overlap that per-symbol measurements cannot fully resolve. A collective measurement instead acts on all $n$ modes jointly as a single quantum system. Even though individual symbols overlap poorly, codewords of $n$ symbols span a joint $n$-mode Hilbert space in which they can be made nearly orthogonal, so the total information extracted exceeds the sum of the $n$ independent single-symbol measurements \cite{Takeoka2004,Guha2011StructuredReceivers}. Saturating the Holevo limit thus requires receivers capable of entangling operations across many received modes~\cite{Guha2011StructuredReceivers,GuhaDuttonShapiro2011JDR,Cui2025GreenMachine}. Squeezed light communications offers a practical alternative without modifying the classical receiver architecture. Redistributing quantum noise via squeezing can raise the effective $\mathrm{SNR}$ and improve capacity beyond the one-quadrature Shannon limit toward the Holevo limit.

The energy efficiency of a communication link, defined as the energy consumed per bit of information, is $E_\mathrm{b} = \hbar\omega\langle N\rangle/C$, where $C$ is the channel capacity. At low $\langle N\rangle$, the Shannon-limited energy efficiency approaches a constant floor $E_{\mathrm{Sh}} \approx \hbar\omega/\log_2(e)$, while the Holevo-limited energy efficiency $E_{\mathrm{Hol}} \approx \hbar\omega/\log_2(e/\langle N\rangle)$ can be made arbitrarily small by reducing $\langle N\rangle$ and spreading information over many modes with collective measurements. This motivates architectures with massively parallelized, low-photon-number channels \cite{GursesInterconnects2026}.

As described in Fig.~\ref{fig:fig1}, a squeezed light transmitter can be designed by generating squeezed vacuum via cascaded SHG and SPDC, followed by passing the squeezed vacuum along with the local oscillator to an interferometer where the squeezed vacuum state is coherently displaced and modulated with a phase modulator to encode data. The squeezed light receiver demonstrated in this work enables sampling of the squeezed quadrature by phase-locking the LO to the minimum-noise axis and projecting the incoming field onto it via coherent detection. This yields sub-shot-noise-limited sensitivity, as demonstrated by the squeezing measurements in Fig.~\ref{fig:fig5}.

From the coherent receiver model, in the high-LO, high-$\mathrm{CMRR}$ limit, the $\mathrm{SNR}$ of a coherent receiver is $\mathrm{SNR} = q_\mathrm{s}^2/\langle(\Delta \hat{q}_\mathrm{s})^2\rangle$, where $q_\mathrm{s}$ is the quadrature displacement amplitude. Throughout this section we consider single-quadrature modulation with displacement only along $q_\mathrm{s}$ and the LO phase-locked to the signal quadrature, so that the mean photon number $\langle N \rangle=q_\mathrm{s}^2$. For a coherent state, $\langle(\Delta \hat{q})^2\rangle = 1/4$, giving $\mathrm{SNR} = 4q_\mathrm{s}^2$. For a displaced squeezed state with squeezing parameter $r$ and the LO phase-locked to the squeezed quadrature, the quadrature variance is reduced to $\langle(\Delta \hat{q})^2\rangle = \frac{1}{4}e^{-2r}$. Including the detection efficiency $\eta$ from \eqref{eq:eta}, which accounts for both optical loss and finite shot noise clearance, the measured quadrature variance is given by \eqref{eq:Veff}. Since $q_\mathrm{s}$ is set by the phase modulator and is independent of $r$, the $\mathrm{SNR}$ for a displaced squeezed state becomes
\begin{align}
\label{eq:SNRsq}
    \mathrm{SNR_{sq}} = \frac{4q_\mathrm{s}^2}{\eta e^{-2r} + 1 - \eta}
\end{align}
At $r=0$, this recovers the coherent-state $\mathrm{SNR}$ $= 4q_\mathrm{s}^2$. Substituting $q_\mathrm{s}^2 = s_\mathrm{s}/(\hbar\omega B)$, where $B$ is the bandwidth,
\begin{align}
\label{eq:SNRsq2}
    \mathrm{SNR_{sq}} = \frac{4s_\mathrm{s}}{\hbar\omega B\left(\eta e^{-2r} + 1 - \eta\right)}
\end{align}
connecting the signal power $s_\mathrm{s}$ and LO power directly to the achievable $\mathrm{SNR}$ through the detection efficiency $\eta$ \eqref{eq:eta}.

For coherent detection on a single quadrature, the channel capacity per mode for a coherent state is the one-quadrature Shannon limit \eqref{eq:CS1}, which in terms of signal power and bandwidth is
\begin{align}
\label{eq:Ccoh}
    C_{\mathrm{S1}} = \frac{1}{2}\log_2\left(1 + \frac{4s_\mathrm{s}}{\hbar\omega B}\right)
\end{align}
where we identified $\langle N\rangle = s_\mathrm{s}/(\hbar\omega B)$. With squeezed light, the reduced noise floor increases the effective $\mathrm{SNR}$, raising the capacity to
\begin{align}
\label{eq:Csq}
    C_{\mathrm{sq}} = \frac{1}{2}\log_2\left(1 + \frac{4s_\mathrm{s}}{\hbar\omega B\left(\eta e^{-2r} + 1 - \eta\right)}\right)
\end{align}
At $\eta = 1$, this simplifies to $C_{\mathrm{sq}} = \frac{1}{2}\log_2(1 + 4\langle N\rangle e^{2r})$, showing that squeezing exponentially enhances the $\mathrm{SNR}$ inside the logarithm. The Holevo limit \eqref{eq:Holevo} remains the ultimate upper bound, giving the hierarchy
\begin{align}
    C_{\mathrm{S1}} \leq C_{\mathrm{sq}} \leq C_{\mathrm{Hol}}
\end{align}
squeezed light communications with coherent detection therefore occupies an intermediate regime, exceeding the one-quadrature Shannon limit without requiring the collective measurements needed to saturate the Holevo limit.

\begin{figure*}[t!]
    \centering
    \includegraphics[width=\linewidth]{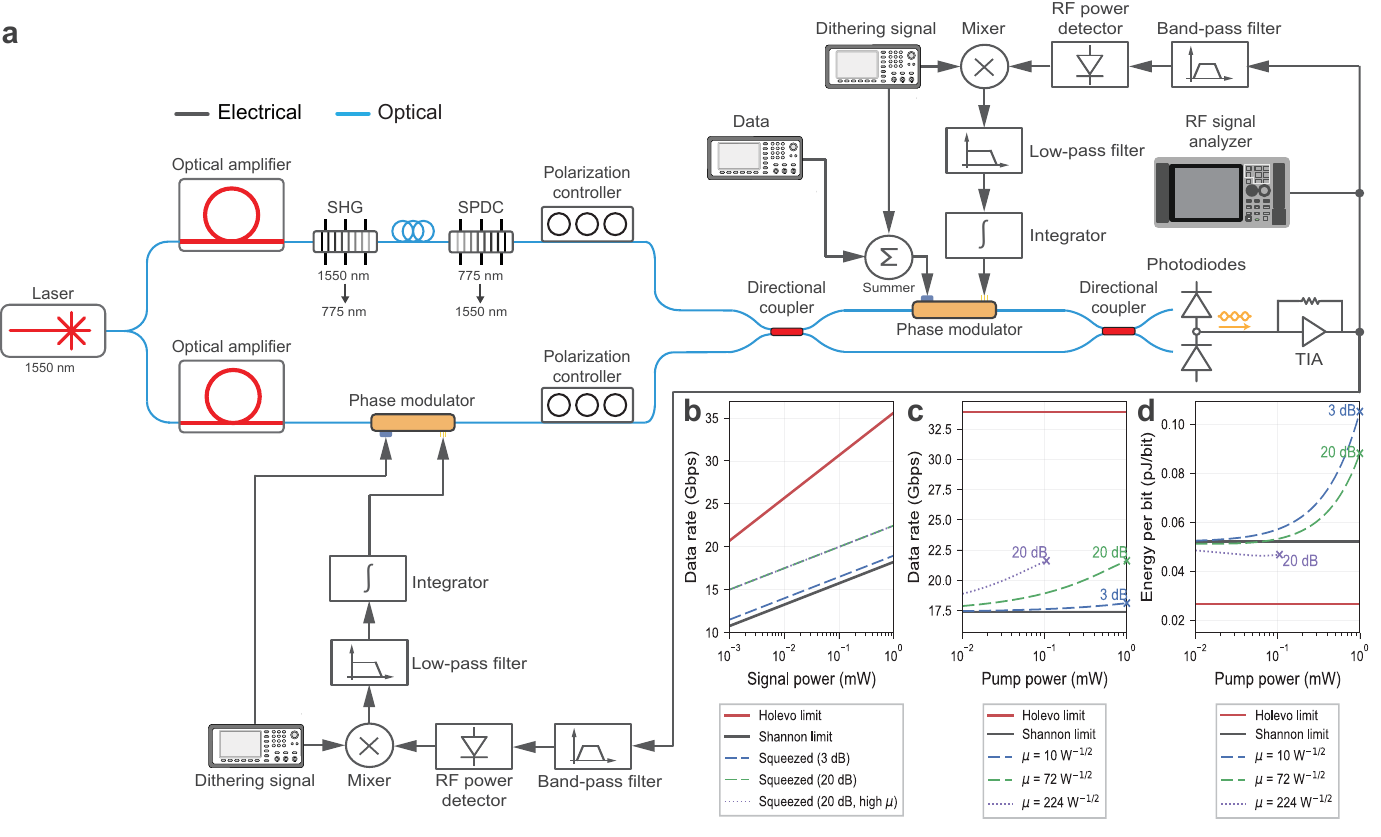}
    \caption{\textbf{Squeezed light communication experiment.} \textbf{a} A 1550 nm continuous-wave laser is split equally into local oscillator (LO) and squeezer paths. Both paths are interfered at an MZI with a phase modulator in one arm driven by the sum of a low-frequency dither tone and the data signal. The LO path has a phase modulator driven by a dither tone alone. The MZI outputs feed a balanced Germanium photodiode pair, and the photocurrent is amplified by a TIA and sent to an RF signal analyzer. Two PLLs act on the two phase modulators, one locks the LO to the squeezed quadrature, the other maintains the MZI balance point for common-mode rejection.
    \textbf{b} Data rate vs. signal power for squeezed light transceivers. Pump power is fixed at 1~\unit{mW} for \unit{\mu} = 10 and 72 \unit{W^{-1/2}} (3 and 20 \unit{dB} squeezing), and at 0.105 \unit{mW} for \unit{\mu} = 224 \unit{W^{-1/2}} (20~\unit{dB} squeezing). Receiver bandwidth is 1.5~\unit{GHz}. Squeezing consistently raises capacity above the Shannon limit toward the Holevo limit.
    \textbf{c,d} Both panels share a fixed signal and LO power of 452.4 \unit{\micro\watt} and receiver bandwidth at 1.5~\unit{GHz}, with squeezing level determined by the pump power and \unit{\mu}. For $\mu =$ 224~\unit{W^{-1/2}}, pump power is limited to 0.105~\unit{mW} by the receiver $\mathrm{SNC}$. Squeezing levels at the cutoff pump power are indicated by markers at the end of each curve. Detection efficiency $\eta=0.99$ assumed throughout.
    \textbf{c} Data rate vs. pump power. Higher \unit{\mu} achieves greater squeezing at a lower pump power, raising capacity above the Shannon limit.
    \textbf{d} Energy per bit vs. pump power. Operating points where curves fall below the Shannon limit (black) simultaneously improve capacity and reduce energy per bit relative to classical coherent detection.}
    \label{fig:fig6}
\end{figure*}

However, the exponential $\mathrm{SNR}$ enhancement comes at the cost of the pump power $P_{\mathrm{pump}}$ required to drive the parametric process. For a nonlinear waveguide with parametric gain coefficient $\mu$, the squeezing parameter $r$ scales as $r =\mu\sqrt{P_{\mathrm{pump}}}$ in the low-conversion limit. The pump power needed to achieve a given $r$ is thus $P_{\mathrm{pump}}=(r/\mu)^2$.
The total optical power consumed by the transceiver is
\begin{align}
\label{eq:Ptotal}
    P_{\mathrm{total}} = s_\mathrm{s} + s_\mathrm{LO} + \left(\frac{r}{\mu}\right)^2
\end{align}
where $s_\mathrm{s}$ is the signal power and $s_\mathrm{LO}$ is the local oscillator power. With squeezing, the energy per bit is
\begin{align}
\label{eq:Eb}
    E_\mathrm{\mathrm{b},\mathrm{sq}} = \frac{P_{\mathrm{total}}}{B \cdot C_{\mathrm{sq}}} = \frac{s_\mathrm{s} + s_\mathrm{LO} + \left(\frac{r}{\mu}\right)^2}{B \cdot \frac{1}{2}\log_2\left(1 + \frac{4s_\mathrm{s}}{\hbar\omega B\left(\eta e^{-2r} + 1 - \eta\right)}\right)}
\end{align}

For the coherent baseline ($r=0$), $P_{\mathrm{total}} = s_\mathrm{s} + s_\mathrm{LO}$ and the energy per bit is

\begin{align}
\label{eq:Ebcoh}
    E_\mathrm{b,coh} = \frac{P_{\mathrm{total}}}{B \cdot C_{\mathrm{coh}}} = \frac{s_\mathrm{s} + s_\mathrm{LO}}{B \cdot \frac{1}{2}\log_2\left(1 + \frac{4s_\mathrm{s}}{\hbar\omega B}\right)}
\end{align}

Squeezing is advantageous when $E_{\mathrm{b},\mathrm{sq}} < E_{\mathrm{b},\mathrm{S1}}$, meaning the $\mathrm{SNR}$ gain from noise suppression $e^{-2r}$ outweighs the pump power overhead $(r/\mu)^2$. A higher parametric gain coefficient $\mu$ lowers this pump power cost for a given squeezing level, enabling the crossover to net squeezing advantage in energy efficiency at lower pump powers.

In \eqref{eq:Eb} the LO power $s_\mathrm{LO}$ contributes directly to the power budget in the numerator, but also sets the detection efficiency $\eta$ through \eqref{eq:eta}, which controls how much of the squeezing advantage is actually preserved in the denominator. Below $P_\mathrm{knee}$, increasing $s_\mathrm{LO}$ rapidly improves $\eta$ and the recoverable squeezing. Beyond $P_{\mathrm{knee}}$, $\eta_{\mathrm{SNC}}$ saturates toward 1 and $\eta$ approaches $\eta_{opt}$, preserving the squeezed noise floor. Further increases in $s_\mathrm{LO}$ add to the power budget without meaningfully improving capacity. There is therefore an optimal $s_\mathrm{LO}$ that balances detection efficiency against power consumption.

To realize a squeezed light communication link, the experimental setup shown in Fig.~\ref{fig:fig6}a extends the squeezed light measurement system with data encoding and phase-locked detection. The signal path generates a displaced squeezed vacuum as described above, where the phase modulator simultaneously encodes data symbols and applies a low-frequency dither tone for phase locking. At the receiver, a phase-locked loop (PLL) extracts the dither tone from the RF output via a bandpass filter and RF power detector, mixes it with the dither reference to generate a DC error signal, and integrates to generate an error signal that stabilizes the LO phase onto the squeezed quadrature. Similarly a second PLL controls the MZI arm phase modulator to maintain the bias point of the interferometer to ensure a balanced output. The data signal rides on top of the dither signal and is at a much higher frequency than the dither frequency. Together the two PLLs maintain quadrature alignment and good common-mode rejection throughout operation, preserving the squeezing advantage.

On-chip integration of the phase-locking electronics alongside the quantum coherent receiver would reduce the loop delay and environmental sensitivity of the tabletop feedback paths, enabling more robust phase tracking. Combined with the demonstrated 14.0 \unit{dB} shot noise clearance, 2.57 GHz bandwidth, and 90.2 \unit{dB} $\mathrm{CMRR}$, the integrated QRX provides a platform capable of resolving the sub-shot-noise quadrature statistics required for squeezed light communications at GHz-class data rates. The 32-channel QRX array further demonstrates the scalability needed for parallelized architectures that can exploit the low-photon-number regime where the quantum advantage is most significant.

The theoretical projections in Figure~\ref{fig:fig6}b-d are evaluated using hardware parameters from a state-of-the-art integrated coherent receiver~\cite{Bruynsteen2021,Tasker2024} with a detection efficiency of $\eta = 0.99$, a receiver bandwidth of $B=1.5$~GHz, and a signal power of $s_\mathrm{s}=452.4$~\unit{\micro\watt} for panels c and d. The full set of link parameters is given in Materials and Methods.

Panel b shows data rate vs.\ signal power, showing that increasing the squeezing level raises capacity above the one-quadrature Shannon limit, approaching the Holevo limit at higher squeezing, particularly at low signal powers. Panels c and d show data rate and energy per bit vs.\ pump power at a fixed signal power. At sufficiently high pump powers, higher-$\mu$ curves both exceed the one-quadrature Shannon capacity (panel c) and reduce the energy-per-bit below the classical limit (panel d), identifying a region of the parameter space where squeezed light communications is advantageous in both bandwidth efficiency and energy consumption.

Approaching the Holevo limit requires progressively higher squeezing levels, controlled by the nonlinear waveguide's parametric gain coefficient $\mu$ and overall detection efficiency $\eta$. High $\mu$ is the primary knob for increasing squeezing and is determined by the conversion efficiency of the nonlinear waveguide. For a fixed $\mu$, the achievable squeezing is then fundamentally limited by the optical loss $1 - \eta$. Thus, high end-to-end efficiency and high shot noise clearance are equally important for closing the gap to the Holevo limit.

\section{Discussion}

The integrated QRX achieves quantum-limited noise performance with 14.0 \unit{dB} shot noise clearance, 3.50 GHz shot-noise-limited bandwidth, and 90.2 \unit{dB} $\mathrm{CMRR}$ in a compact photonic-electronic package. At this $\mathrm{CMRR}$, the LO noise contribution to the output is suppressed by more than nine orders of magnitude relative to the signal shot noise, ensuring that the photocurrent faithfully reproduces the quadrature statistics of the input field. This is the essential requirement for squeezed light communications, where sub-shot-noise sensitivity must be maintained at GHz rates.

The 32-channel array extends this performance to scale, with a median $\mathrm{SNC}$ of 26.6 \unit{dB} and $P_\mathrm{knee}$ of 12.6 $\mathrm{\mu W}$. The low $P_\mathrm{knee}$ is significant for parallelized Holevo-limited architectures: since the improvement factor $F \approx 1 - \ln \langle N\rangle$ grows at low mean photon number per channel, a practical strategy distributes a fixed photon budget across many channels each operating at low $\langle N\rangle$. The demonstrated array can scale to hundreds or thousands of channels before reaching the two-photon-absorption ceiling in silicon waveguides \cite{Chung2018}, with each channel entering the shot-noise-limited regime at microwatt-level LO power.

The observed squeezing of $0.15 \pm 0.01$ \unit{dB} is limited by the fiber-optic source and tabletop losses, not the on-chip receiver. With 2.7 \unit{dB} total on-chip loss and $\mathrm{SNC}$ exceeding 10 \unit{dB} up to 2.24 GHz, the receiver can support up to 3 \unit{dB} of observable squeezing if paired with a lower-loss source. Reducing edge coupler loss and improving photodiode quantum efficiency would directly deepen observable squeezing and increase the capacity enhancement in a communication link.

The proposed dual-PLL communication experiment addresses the two critical phase stability requirements: tracking the LO to the squeezed quadrature and maintaining balanced detection for $\mathrm{CMRR}$. Any residual phase error $\delta\phi$ degrades the effective noise variance to $V_{\mathrm{eff}} = \frac{1}{4}(e^{-2r}\cos^2\delta\phi + e^{2r}\sin^2\delta\phi)$, but for accessible squeezing levels ($r \approx 0.1$--$0.5$), maintaining $\delta\phi < 5^{\circ}$ preserves more than 90\% of the squeezing advantage, well within the demonstrated PLL and $\mathrm{CMRR}$ stability.

Squeezed states with coherent detection do not saturate the Holevo limit, which requires collective measurements over many channel uses, but they represent a practical intermediate step that exceeds the one-quadrature Shannon limit. The path to fully Holevo-limited communications requires reconfigurable multimode interferometers and non-Gaussian operations at the receiver \cite{GursesInterconnects2026,Cui2025GreenMachine}. The coherent receiver demonstrated here on a silicon photonics platform, with its dense waveguide routing, scalable photodetection, and robust phase coherence, provides a natural foundation for integrating such interferometric layers. The quantum coherent transceiver platform thus bridges conventional coherent communications and the quantum-limited regime, providing the essential building blocks for squeezed light communications and, ultimately, quantum interconnects approaching the Holevo limit.

\section{Materials and Methods}

\subsection{Theory background}
We can model coherent detection through two viewpoints: semi-classical and quantum. In the semi-classical treatment, we quantize the optical power as a variable and link this quantization to the discretization in photon number. In the quantum treatment, we account for the second quantization of the electromagnetic field with the creation-annihilation operator formalism \cite{gerry2005introductory}.
\subsubsection{Semi-classical treatment}
In this treatment, a single-mode field in a cavity can be defined with the following electric and magnetic fields:
\begin{align} \begin{gathered}
    E_\mathrm{x}(z,t)=C X(t)\sin{(kz)}\\
    B_\mathrm{y}(z,t)=C\left(\frac{\mu_0\epsilon_0}{k}\right)P(t)\cos{(kz)}
\end{gathered} \end{align}
where $C=\left(\frac{2\omega^2}{V\varepsilon_0}\right)^{1/2}$, $V$ is the mode volume, $k$ is the wavenumber, $\omega$ is the angular frequency, and $\mu_0, \epsilon_0$ are the vacuum permeability and permittivity, respectively. $X(t)=X_0\sin(\omega t)$ and $P(t)=\omega X_0\cos(\omega t)$ are the canonical position and momentum, respectively, satisfying the wave equation. The Hamiltonian signifying the field energy is of the same form as a harmonic oscillator with unit mass.
\begin{align} \begin{aligned}
    H&=\frac{1}{2}\int dV\left[\epsilon_0E_\mathrm{x}^2(z,t)+\frac{1}{\mu_0}B_\mathrm{y}^2(z,t)\right]\\
    &= \frac{1}{2}(P^2 +\omega^2 X^2)
    \label{eq:Hamiltonian}
\end{aligned} \end{align}
Assuming perfect coupling to a medium where the field propagates as a traveling wave, for which the electric and magnetic field are now in phase in both space and time, we can represent the electric and magnetic fields as phasors with complex amplitudes, $\varepsilon_0$ and $\beta_0$.
\begin{align}
\begin{gathered}
    E_\mathrm{x}(z,t)=\varepsilon_0e^{i(kz-\omega t)}+\mathrm{c.c.}\\
    B_\mathrm{y}(z,t)=\beta_0e^{i(kz-\omega t)}+\mathrm{c.c.}
\end{gathered} \end{align}
where c.c. is the complex conjugate, and for a plane wave, $\beta_0=\varepsilon_0 c$, where $c=\frac{1}{\sqrt{\mu_0\epsilon_0}}$ is the speed of light. We can derive the time-averaged Poynting vector, $\langle \textbf{S}(z,t) \rangle$, of the field escaping from the cavity,
\begin{align} \begin{aligned}
    \langle S(z,t)\rangle&=\frac{1}{2}\mathrm{Re}\left\{ E_\mathrm{x}(z,t)\times \frac{1}{\mu_0}B_\mathrm{y}^*(z,t)\right\}=\frac{1}{2}\epsilon_0c\left|\varepsilon_0\right|^2
\end{aligned} \end{align}
Multiplying this over the area of interest, we can derive the power of this classical electromagnetic field.
\begin{align} \begin{aligned}
\label{eq:S}
    S=\frac{1}{2}\epsilon_0c\left|\varepsilon_0\right|^2A
\end{aligned} \end{align}
Now, to account for noise, we redefine $S$ to be a random variable with signal-carrying, $s$, and noise-carrying, $\Delta s$ components.
\begin{align} \begin{aligned}
\label{eq:Ss}
    S=s+\Delta s
\end{aligned} \end{align}
Assuming discretization of this power to units of photons with a photon energy of $\hbar \omega$, $S=\frac{\hbar\omega}{T}N$, where $N$ is the photon number over a measurement period of $T$. Then, we can define $s=\langle S\rangle=\frac{\hbar\omega}{T}\langle N\rangle$ as mean power and $\Delta s=S-\langle S\rangle$ as power fluctuations.

For coherent states of light, photon number statistics follow a Poissonian distribution allowing us to define $\langle \Delta s^2\rangle=(\frac{\hbar\omega}{T}\sqrt{\langle N\rangle})^2=\frac{\hbar\omega}{T} s$. Therefore, with direct detection, the signal-to-noise ratio ($\mathrm{SNR}$) for coherent states is
\begin{align}
    \mathrm{SNR_{dd}}=\frac{s^2}{\langle \Delta s^2\rangle}=\frac{s}{\hbar\omega/T}
\end{align}

\subsubsection{Quantum treatment}
In this treatment, we quantize the harmonic oscillator with the Hamiltonian of form \eqref{eq:Hamiltonian} by taking $X \rightarrow \hat{X}$ and $P \rightarrow \hat{P}$, where the canonical position and momentum operators are Hermitian and observables. The Hamiltonian operator is then
\begin{align} \begin{aligned}
    \hat{H}=\frac{1}{2}(\hat{P}^2+\omega^2\hat{X}^2)
\end{aligned} \end{align}
Solving the time-independent Schrodinger's equation, $\hat{H}\ket{\psi}=E\ket{\psi}$, gives the energy levels of the quantum harmonic oscillator.
\begin{align} \begin{aligned}
    E=\hbar\omega\left( n+\frac{1}{2}\right)
\end{aligned} \end{align}
implying $\hat{H}=\hbar\omega\left(\hat{N}+\frac{1}{2}\right)$, where $\hat{N}$ is the photon number operator satisfying $\hat{N}\ket{\psi}=n\ket{\psi}$. Now, let $\hat{a}$ be a non-Hermitian operator that satisfies
\begin{align} \begin{aligned}
    \hat{N}=\hat{a}^\dagger\hat{a}
\end{aligned} \end{align}
where $\left[\hat{a},\hat{a}^\dagger\right]=c$, making $\hat{a}\hat{a}^\dagger=\hat{N}+c$. Setting $c=1$, the Hamiltonian is $\hat{H}=\hbar\omega\frac{\hat{a}^\dagger\hat{a}+\hat{a}\hat{a}^\dagger}{2}$. With this, we can define $\hat{a}$ and $\hat{a}^\dagger$ in terms of $\hat{X}$ and $\hat{P}$.
\begin{align} \begin{aligned}
    \begin{gathered}
    \hat{a} = (2\hbar \omega)^{-1/2}(\omega \hat{X}+i\hat{P})\\
    \hat{a}^\dagger = (2\hbar \omega)^{-1/2}(\omega \hat{X}-i\hat{P})
    \end{gathered}
\end{aligned} \end{align}
Since,
\begin{align} \begin{aligned}
        \hat{a}\ket{\psi_{n}}&=\sqrt{n}\ket{\psi_{n-1}}\\
        \hat{a}^\dagger\ket{\psi_{n}}&=\sqrt{n+1}\ket{\psi_{n+1}}
\end{aligned} \end{align}
these operators are the creation and annihilation operators that correspond to the creation and destruction of single photon in the quantum harmonic oscillator description of the field mode. In this formalism, the corresponding electric and magnetic field operators are
\begin{align} \begin{aligned}\begin{aligned}
    \hat{E}_\mathrm{x}(z,t) &= E_0 (\hat{a}+\hat{a}^\dagger)\sin{kz}\\
    \hat{B}_\mathrm{y}(z,t) &= B_0 \frac{1}{i}(\hat{a}-\hat{a}^\dagger)\cos{kz}
\end{aligned}\end{aligned} \end{align}
Now, in this treatment, light with a field operator $\hat{a}$ can again be defined with signal-carrying and noise-carrying components.
\begin{align} \begin{aligned}
    \hat{a}&=a+\Delta \hat{a}
\end{aligned} \end{align}
where $a=\langle \hat{a} \rangle$ is the mean and $\Delta \hat{a}=\hat{a}-\langle \hat{a} \rangle$ is the fluctuation term, which is also an operator.
We can further refactor the field into two quadrature operators that correspond to the canonical position and momentum operators presented above,
\begin{align}
\begin{aligned}
    \begin{gathered}
\hat{q}=\frac{\hat{a}+\hat{a}^\dagger}{2}=\sqrt{\frac{\omega}{2\hbar}}\hat{X},\\ \hat{p}=\frac{\hat{a}-\hat{a}^\dagger}{2i}=\sqrt{\frac{1}{2\hbar\omega}}\hat{P}.
    \end{gathered}
\end{aligned}
\end{align}
In phase space, we can separate the field into two quadrature operators that correspond to the canonical position and momentum operators from earlier.
\begin{align} \begin{aligned}
\label{eq:a_q_p}
    \hat{a}&=\hat{q}+i\hat{p}=q+\Delta \hat{q}+i(p+\Delta \hat{p})
\end{aligned} \end{align}
now where $a=q+ip$, $\Delta \hat{a}=\Delta \hat{q}+i\Delta \hat{p}$ and $\hat{q}=q+\Delta \hat{q}$, $\hat{p}=p+\Delta \hat{p}$. We can set $p=0$ by choosing our frame of reference to align with one quadrature. Then,
\begin{align} \begin{aligned}
    \hat{a}&=q+\Delta \hat{q}+i\Delta \hat{p}
    \label{eq:nsq}
\end{aligned} \end{align}
The commutation relations are $[\hat{q},\hat{q}]=[\hat{p},\hat{p}]=0$, $[\hat{q},\hat{p}]=i/2$, $\left[\Delta \hat{q},\Delta\hat{p}\right]=i/2$.

To unite both treatments, we assume a stationary field with ensemble measurements taken over a period of $T$ time. With this assumption, we can define an effective power operator same as the power in the semiclassical treatment in terms of the photon number operator.
\begin{align} \begin{aligned}
\label{eq:power}
    \hat{S}&=\frac{\hbar\omega}{T}\hat{N}=\frac{\hbar\omega}{T}(n+\Delta \hat{n})=s+\Delta \hat{s}\\
    &=\frac{\hbar\omega}{T}(\hat{a}^\dagger\hat{a})=\frac{\hbar\omega}{T}\left(q^2+2q\Delta \hat{q}+\Delta \hat{p}^2 + \Delta \hat{q}^2 - \frac{1}{2}\right)
\end{aligned} \end{align}
where $s=\langle\hat{S}\rangle$ is the mean power and $\Delta \hat{s}=\hat{S}-\langle\hat{S}\rangle$ is the power fluctuations. The signal power in terms of quadrature operators is
\begin{align} \begin{aligned}
    s^2&=\left(\frac{\hbar\omega}{T}\right)^2\langle\hat{a}^\dagger \hat{a}\rangle^2\\
    &=\left(\frac{\hbar\omega}{T}\right)^2\left[q^2+\langle (\Delta \hat{q})^2\rangle+\langle (\Delta \hat{p})^2\rangle-\frac{1}{2}\right]^2
\end{aligned} \end{align}
where we used the fact that $\hat{q}, \hat{p}$ are Hermitian and $\left[\Delta \hat{q},\Delta\hat{p}\right]=i/2$. For coherent states, $\langle(\Delta \hat{q})^2\rangle=\langle(\Delta \hat{p})^2\rangle=\frac{1}{4}$, making $s^2=\left(\frac{\hbar\omega}{T}\right)^2 q^4$. The power fluctuation in terms of quadrature operators is
\begin{align} \begin{aligned}
   \Delta \hat{s}=\frac{\hbar\omega}{T}(\hat{a}^\dagger \hat{a}-\langle \hat{a}^\dagger \hat{a}\rangle)
   =\frac{\hbar\omega}{T}\left\{2q\Delta \hat{q}+\delta \hat{q}+\delta \hat{p}\right\}
\end{aligned} \end{align}
where $\delta \hat{q}=\left[(\Delta \hat{q})^2-\langle(\Delta \hat{q})^2\rangle\right]$ and $\delta\hat{p}=\left[(\Delta \hat{p})^2-\langle(\Delta \hat{p})^2\rangle\right]$. With this, the noise power in $\mathrm{SNR}$ is
\begin{align} \begin{aligned}
    \langle \Delta \hat{s}^2\rangle&=\left(\frac{\hbar\omega}{T}\right)^2\Big[4q^2\langle (\Delta \hat{q})^2 \rangle
+ \langle\delta \hat{q}^2\rangle
+ \langle\delta \hat{p}^2\rangle\\
&+ 4q\,\left[ \langle (\Delta \hat{q})^3 \rangle + \langle \Delta \hat{q}\,(\Delta \hat{p})^2 \rangle \right]
+ 2\mathrm{Cov}(\Delta \hat{q}^2,\Delta \hat{p}^2)\Big]
\end{aligned} \end{align}
For Gaussian states, third-order terms are zero due to the symmetry in the phase space and $\langle\delta \hat{q}^2\rangle=2\langle(\Delta\hat{q})^2\rangle^2$, $\langle\delta \hat{p}^2\rangle=2\langle(\Delta\hat{p})^2\rangle^2$. The covariance between the quadrature power fluctuations is $\mathrm{Cov}(\Delta \hat{q}^2,\Delta \hat{p}^2)=C_\mathrm{sym}-\frac{1}{8}$, where $C_\mathrm{sym}$ is the state-dependent contribution from the symmetrically ordered moments and $-\frac{1}{8}$ is a universal correction from the commutator $\left[\Delta \hat{q},\Delta\hat{p}\right]=i/2$. For Gaussian states, $C_\mathrm{sym}=0$, giving $\mathrm{Cov}(\Delta \hat{q}^2,\Delta \hat{p}^2)=-\frac{1}{8}$. Restricting to Gaussian states, which include all states considered in this work,
\begin{align} \begin{aligned}
    \langle \Delta \hat{s}^2\rangle&=\left(\frac{\hbar\omega}{T}\right)^2\\
    &\left[4q^2\langle (\Delta \hat{q})^2 \rangle
+ 2\langle(\Delta\hat{q})^2\rangle^2 + 2\langle(\Delta\hat{p})^2\rangle^2 - \frac{1}{4}\right]
\end{aligned} \end{align}
For coherent states, the noise power is $\langle \Delta \hat{s}^2\rangle=(\hbar\omega/T)^2 q^2$. Then, the $\mathrm{SNR}$ for direct detection is
\begin{align}
    \mathrm{SNR_{dd}}=\frac{s^2}{\langle\Delta s^2\rangle}=q^2=\frac{s}{\hbar\omega/T}
\end{align}
exactly matching the semi-classical treatment and uniting both approaches.

\subsection{Chip design and fabrication}
The photonic integrated circuit (PIC) section of the QRX comprises two edge couplers with 127 $\mathrm{\mu m}$ pitch suitable for coupling the local oscillator (LO) and signal light with fiber arrays, a thermo-optic phase shifter (TOPS) for modulating the phase, a push-pull Mach-Zehnder interferometer (MZI) for interfering LO and signal and correcting the common-mode rejection ratio ($\mathrm{CMRR}$), and a balanced pair of Ge photodiodes (PDs) for photodetection. The electronic integrated circuit (EIC) section of the QRX comprises a transimpedance amplifier (TIA), a voltage amplifier, and an output buffer with differential outputs. The EIC, which is a bare die (ONET4291T), has a transimpedance gain of 3.2 kΩ and a 3-dB bandwidth of 2.8 GHz \cite{onet4291t}. As seen in Fig.~\ref{fig:fig3}a, the PIC and EIC are packaged together on a custom PCB with a wirebond between the PIC RF output and EIC RF input. To minimize the bond wire parasitics before EIC amplification, pads on both chips were kept level and positioned as close together as possible. RF outputs from the EIC were wirebonded to transmission lines on the PCB matched to 50 $\mathrm{\Omega}$. The transmission lines were connected to SMA connectors on the PCB to read out the differential RF outputs. All other DC lines were also wirebonded to the PCB. The PIC was fabricated with Advanced Micro Foundry using a 193 nm silicon-on-insulator (SOI) process. The process has two metal layers (2000-nm thick and 750-nm thick) for electronic routing, a titanium nitride heater layer, a 220-nm thick silicon layer, a 400-nm thick silicon nitride layer, germanium epitaxy, and various implantations for active devices. A process design kit (PDK) from the foundry was provided. The PIC was laid out using KLayout and Cadence Virtoso and was simulated using Lumerical for design verification.\par

\subsection{Coherent receiver characterization}
\subsubsection{Bandwidth and frequency response}
The packaged QRX was first characterized to determine its bandwidth, $\mathrm{CMRR}$, and shot noise clearance to ensure quantum-limited performance. Bandwidth and $\mathrm{CMRR}$ measurements were done as a three-port measurement with a 20 GHz vector network analyzer (VNA) connected to the differential outputs of the EIC and a 40 GHz amplitude modulator. Current meters were connected to the DC bias lines of the PDs to monitor the PD photocurrents. The optoelectronic response of the QRX is shown in Fig.~\ref{fig:fig3}d, showing a 3-\unit{dB} bandwidth of 2.57 GHz.

\subsubsection{Shot noise clearance}
Shot noise clearance measurements were done with an RF spectrum analyzer (ESA) by sweeping the LO power and measuring the RF spectrum of one of the differential outputs. The photocurrents were monitored to maximize the $\mathrm{CMRR}$ and minimize the LO shot noise and relative intensity noise from leaking in \cite{Gurses2022}. The measurements were also done in a Faraday cage to minimize RF interference from external sources. As seen in Fig.~\ref{fig:fig3}b and ~\ref{fig:fig3}c, the shot noise clearance is 14.0 \unit{dB} over the bandwidth of the QRX. The $\mathrm{SNC}$ frequency response with maximum LO photocurrent is seen in Fig.~\ref{fig:fig3}e, showing a shot-noise-limited bandwidth of 3.50 GHz. A shot noise line is also fitted to the electronic noise subtracted measurements showing a near-unity slope of 1.007 $\mathrm{\pm}$ 0.015, ensuring broadband quantum-limited performance.

\subsubsection{Common-mode rejection ratio}
$\mathrm{CMRR}$ is measured by setting the push-pull MZI in the PIC to unbalanced (100:0) and balanced (50:50) settings while injecting intensity-modulated LO to the QRX. A 1550 nm source (APEX AP3350A) is intensity modulated at 1.1 MHz, and the modulated light is sent to the LO port of the QRX. The QRX output is connected to an RF signal analyzer to measure the RF power at 1.1 MHz. Two measurements are taken by tuning the MZI to unbalanced and balanced settings. Following the definition in \eqref{eq:CMRR1} and staying consistent with literature \cite{Tasker2021,Bruynsteen2021,Gurses2022,Gurses2024}, we define the experimental $\mathrm{CMRR}$ as the ratio between the RF power of a single photodiode and the RF power when the photodiode currents are subtracted, calculated as

\begin{align}
\label{eq:CMRR2}
\mathrm{CMRR}=10\log_{10}\left(\frac{P_\mathrm{unb}}{4P_\mathrm{bal}}\right)
\end{align}
where $P_\mathrm{unb}$ and $P_\mathrm{bal}$ are the unbalanced and balanced RF power measurements, respectively. The $\mathrm{CMRR}$ measurements were taken with the signal analyzer using a frequency span of 10 kHz, resolution and video bandwidths of 100 Hz, and a center frequency of 1.1 MHz. To validate the stability of $\mathrm{CMRR}$, ten snapshots were taken over the course of 10 seconds for both balanced and unbalanced settings. The $\mathrm{CMRR}$ was then calculated from the average of the traces in these snapshots. In the initial measurement, due to the limited dynamic range of the ESA, the balanced measurement was able to suppress the 1.1 MHz peak below the noise floor as seen in Fig.~\ref{fig:fig3}g. The average $\mathrm{CMRR}$ from 10 traces measured over 10 seconds from this initial measurement is 80.6 \unit{dB}, calculated from the ratio between the balanced and unbalanced measurements as $\mathrm{CMRR=86.6\,dB-2\times3\,dB=80.6\,dB}$. To increase the dynamic range and resolve the suppressed peak, the LO optical power in balanced measurement was increased by 18.2 \unit{dB}. The LO optical power difference between balanced and unbalanced measurements was monitored and recorded throughout the measurements with an average power difference of 18.2 \unit{dB}. Since the change in RF power has a quadratic relation with the change in optical power, we add double the optical power difference in \unit{dB} to the measured peak amplitude difference. The resulting traces after changing the LO power in the balanced measurement are shown in Fig.~\ref{fig:fig3}h. The average $\mathrm{CMRR}$ from 10 traces measured over 10 seconds from this final measurement is 90.2 \unit{dB} ($\mathrm{CMRR=59.8\,dB+2\times18.2\,dB-2\times3\,dB=90.2\,dB}$), with a maximum $\mathrm{CMRR}$ of 92.3 dB ($\mathrm{CMRR=61.9\,dB+2\times18.2\,dB-2\times3\,dB=92.3\,dB}$).

\subsection{32-channel array characterization}
\subsubsection{32-channel CMRR}
To quantify the effectiveness of the $\mathrm{CMRR}$ auto-correction across the array, we injected an intensity-modulated LO into the PIC and recorded 10 ms time traces from all 32 channels with the feedback loop turned off and on while sweeping the modulation frequency from 1 kHz to 50 MHz. The traces were converted to the frequency domain, and the change in the modulation-tone amplitude with auto-correction enabled was used to quantify the $\mathrm{CMRR}$ improvement of each channel, which provides a lower bound on the true channel $\mathrm{CMRR}$.

\subsubsection{32-channel SNC}
We measured the shot-noise clearance of all 32 channels by comparing the noise spectra with and without LO illumination, using a total LO power of 49.3 mW, corresponding to an average of 1.54 mW per channel and set by the onset of photodiode breakdown in the most power-sensitive channel.

\subsection{Squeezed light measurement setup}
To demonstrate squeezed light detection across a high bandwidth, a transceiver was assembled in combination with the integrated QRX and a fiber-optic QTX. The QTX includes a 1550 nm laser source that splits the light into two paths, one for the signal path to generate squeezed vacuum and one for the LO path. In the LO path, there is an optical amplifier, a phase modulator that controls the LO phase, and a polarization controller. In the signal path, there is an optical amplifier, two periodically-poled lithium niobate (PPLN) waveguides cascaded with reverse polarities for SHG and SPDC, an optical switch, an isolator, and a polarization controller. Signal and LO paths are coupled to chip with a V-groove fiber array. Noise floor oscillations in the output with 1 Hz LO phase modulation were measured with a signal analyzer at different sideband frequencies.

Squeezed vacuum is injected into the QRX as signal and coherent light, phase modulated with a LN phase modulator (EOSpace) at 1 Hz, with a continuous-wave power of 9.50 mW is injected into the high-BW QRX as LO. This LO power corresponds to an $\mathrm{SNC}$ of 12.7 \unit{dB} seen in Fig.~\ref{fig:fig3}b. QRX output is connected to an RF signal analyzer (Keysight N9030B). The signal analyzer is used in zero-span mode with 8 MHz resolution bandwidth, 100 kHz video bandwidth, and varying center frequencies up to $\mathrm{BW_{shot}}$.

\subsubsection{Noise level estimation}
The estimation procedure for the noise levels is as follows. For the time-domain trace at each frequency, same number of points are randomly sampled. From these samples, the histograms are constructed for the sampled noise powers, and the PDFs are approximated by kernel density estimation (KDE). The squeezed and anti-squeezed noise level estimates are then obtained from the peaks in the derivative of the KDEs \cite{Gurses2025OnChipPhasedArray}.

\subsection{Squeezed light communication link parameters}
The theoretical projections in Figure~\ref{fig:fig6}b-d are evaluated using hardware parameters from a state-of-the-art integrated coherent receiver~\cite{Bruynsteen2021,Tasker2024}. In particular, we set a receiver bandwidth $B=$1.5~GHz and a knee power of $P_\mathrm{knee}=$4.52~\unit{\micro\watt}. The LO power is set to $s_\mathrm{LO}=$452~\unit{\micro\watt}, corresponding to a shot noise clearance of 20~\unit{dB}. This yields an overall detection efficiency from~\eqref{eq:eta} of $\eta =\eta_{opt}(1-P_\mathrm{knee}/s_\mathrm{LO})=0.99$ with $\eta_{opt}$ assumed to be 1. This choice of $\mathrm{SNC}$ caps the observable squeezing at 20~\unit{dB} while maintaining a high efficiency consistent with literature. The signal power is fixed at $s_\mathrm{s}=s_\mathrm{LO}= $452.4 \unit{\micro\watt}, a choice representative of short-reach datacom receivers, for panels c and d. Conversion efficiencies up to 5,000,000 \%~\unit{W^{-1}} corresponding to a $\mu \approx$ 224~\unit{W^{-1/2}} have been demonstrated in periodically poled lithium niobate microring resonators \cite{Lu2020PPLN}, suggesting that high squeezing levels are within reach.

\bibliography{references.bib}

\end{document}